# A high-resolution hindcast of wind and waves for The North Sea, The Norwegian Sea and The Barents Sea[$]

Magnar Reistad[*], Øyvind Breivik[*,1], Hilde Haakenstad[*], Ole Johan Aarnes[*], Birgitte R. Furevik[*] and Jean-Raymond Bidlot[§]

[1]Corresponding author. E-mail oyvind.breivik@met.no.
[*]Norwegian Meteorological Institute, Norway
[§]European Centre for Medium-Range Weather Forecasts, Reading, United Kingdom

## *Abstract*

A combined high-resolution atmospheric downscaling and wave hindcast based on the ERA-40 reanalysis covering the Norwegian Sea, the North Sea and the Barents Sea is presented. The period covered is from September 1957 to August 2002. The dynamic atmospheric downscaling is performed as a series of short prognostic runs initialized from a blend of ERA-40 and the previous prognostic run to preserve the fine-scale surface features from the high-resolution model while maintaining the large-scale synoptic field from ERA-40. The nested WAM wave model hindcast consists of a coarse 50 km model covering the North Atlantic forced with ERA-40 winds and a nested 10-11 km resolution model forced with downscaled winds.

A comparison against *in situ* and satellite observations of wind and sea state reveals significant improvement in mean values and upper percentiles of wind vectors and the significant wave height over ERA-40. Improvement is also found in the mean wave period. ERA-40 is biased low in wind speed and significant wave height, a bias which is not reproduced by the downscaling. The atmospheric downscaling also reproduces polar lows, which can not be resolved by ERA-40, but the lows are too weak and short-lived as the downscaling is not capable of capturing their full life cycle.

# 1 Introduction

Reliable time series of wind and wave conditions are crucial for establishing climatology, analyzing trends and estimating return values. The design of offshore installations and the planning of offshore operations depend crucially on reliable statistics of wind and waves. However, often there are not sufficient measurements to obtain reliable estimates of the probability distribution of wind and wave parameters. Similarly, the measured time series are usually too short to make realistic simulations of offshore operations, *e.g.* calculate the long-term probability of weather windows and threshold levels of wind and waves. Calculation of long return periods such as 100-year return values requires very long stable time series, see *e.g. Lopatoukhin et al.* [2000].

An atmospheric reanalysis is a rerun of the past using a subset of available observations and a fixed model setup and data assimilation scheme to minimize the error drift (*i.e.* keep the error statistics as stationary as possible by keeping the set of observations as stable as possible) over the period [*Bromwich et al.*, 2007]. The most commonly used global atmospheric reanalyses today are the European Reanalysis project [ERA-40, see *Uppala et al.*, 2005], the 40-year reanalysis of the National Centers for Environmental Prediction/National Center for Atmospheric Research (NCEP/NCAR) reanalysis (NRA), see *Kalnay et al.* [1996] and the Japanese Reanalysis [JRA-25, see *Onogi et al.*, 2005]. The resolution of the ERA-40 and JRA-25 is approximately 125 km, while the resolution of the NCEP/NCAR reanalysis is approximately 210 km. Although these reanalyses represent the best available long-term global statistics, none of these resolves mesoscale dynamics nor can they come close to modeling coastal wind and wave conditions.

Wave hindcast integrations are similar to atmospheric reanalyses in their attempt to recreate past conditions as accurately as possible. However, while atmospheric reanalyses rely on the assimilation of observations, the general scarcity of wave observations, especially before the advent of satellite altimetry, forces wave hindcast integrations to rely on wind forcing without data assimilation. One notable exception is the ERA-40 [*Uppala et al.*, 2005], which contains a wave model coupled to the atmospheric component. These wave hindcast studies generally involved a one-way forcing of the sea state by the 10 m wind fields estimated from wind and pressure observations and atmospheric analyzes [*Haug and Guddal*, 1981; *Eide et al;* 1985; *Reistad and Iden*, 1998; *Cox et al.*, 1995; *WASA-group*, 1998; *Günther et al.*, 1998; *Wang and Swail;* 2001; *Wang and Swail*, 2002; Reistad *et al.*, 2007]. These methods usually yield good results in open-ocean conditions, but the quality of their wind field estimates is limited by the amount of surface observations and coarse resolution. Recently, several wave hindcast studies have employed the wind fields from full atmospheric reanalyses. For example, *Swail and Cox* [2000] and *Cox and Swail* [2001] ran hindcast integrations forced with NRA winds. Similarly, *Sterl et al.* [1998] performed a global wave hindcast study based on the 15-year atmospheric reanalysis ERA-15, the precursor to ERA-40 [*Gibson et al.*, 1996, 1997]. More recently, *Caires and Sterl* [2005] studied the 100-year return estimates of significant wave height based on ERA-40. In European waters, the 44-year HIPOCAS wind database, based on a downscaling of NRA over the period 1958-2001, has spawned several regional wave hindcast studies on roughly 50 km resolution [*Garcia-Sotillo et al.,* 2005*; Ratsimandresy et al.,* 2008*; Pilar et al.*, 2008]. Although the quality of wave hindcasts forced with reanalyzed winds is certainly higher than earlier studies using kinematic estimates of surface winds, the spatial resolution remains poor. Few attempts have been made to date to produce long-term wave hindcasts on high resolution with the notable exception of the Southern North Sea HIPOCAS (5.5 km) wave hindcast [*Weisse et al,* 2002; *Weisse and Günther*, 2007] which was based on a regional downscaling of NRA. A further nesting was carried out for a small region in the German Bight on 400x400 m resolution by *Gaslikova and Weisse* [2006]. For an overview of recent hindcast studies, see *Weisse and von Storch* [2010].

Atmospheric reanalyses and wave hindcasts properly compared against reliable measurements represent powerful proxies for observations in regions and periods where such observations are scarce. But by delivering gap-free time series and area-wide coverage reanalyses and hindcast archives offer much more than a simple substitute for observations. The objective of this study is to assess the quality of the wind and wave statistics of the downscaling/hindcast against *in situ* and satellite observations. By dynamically downscaling ERA-40 atmospheric reanalyses using a high-resolution numerical weather prediction model we aim to resolve the interaction of the large-scale dynamics with regional physiographic details and thus form realistic mesoscale features at approximately 10 km grid resolution. A particular challenge on this scale is the modeling of polar lows. These mesoscale phenomena are difficult to predict due to their relatively small scale, complex model physics, and the general paucity of observations in the Arctic. *Condron et al*. [2005] studied the presence of polar mesocyclones in ERA-40 using an automated cyclone detection algorithm and found that 80 % of cyclones larger than 500 km were captured by ERA-40, but the hit rate (the chance of finding an observed cyclone of a given dimension in the ERA-40 fields) decreased approximately linearly to ~40% for 250 km and to ~20% for 100 km scale cyclones. This is consistent with the observation that features in a spectral model must be 2-4 times the smallest scale to be accurately represented [*Pielke*, 1991; *Laprise*, 1992]. In line with this we do not expect ERA-40 with its spectral truncation limit of T159 (approximately 125 km) to give an accurate representation of features smaller than 500 km, hence polar lows will generally not be present in ERA-40. We will assess to which extent it is possible to detect and model polar lows using the present downscaling approach.

This paper is organized as follows. First, Section 2 gives a general overview of the model setup and the downscaling technique. In Section 3 the results of the atmospheric downscaling and the wave model integration are presented and compared with observations, always with ERA-40 as a baseline reference. Finally, a discussion of the results and to what extent the dynamical downscaling improves on the results of ERA-40 is given in Section 4 before we summarize and conclude in Section 5.

## 2 Downscaling and hindcast configuration

The 45-year atmospheric downscaling of ERA-40 and the wave hindcast are presented separately as there is no feedback from the wave model to the numerical weather prediction model.

### 2.1 Downscaling the ERA-40 reanalysis

ERA-40 is a global coupled atmosphere-wave reanalysis covering the 45 years from September 1957 to August 2002. ERA-40 was produced by The European Centre for Medium-Range Weather Forecasts (ECMWF) using the Integrated Forecasting System (IFS) version CY31R2 with a spectral resolution of T159L60. Analyzed atmospheric fields are available every six hours with approximately 1.125º horizontal resolution. ERA-40 wave fields were generated with the Wave prediction model (WAM), see *WAMDI* [1988], coupled to the atmospheric model and run in deep-water mode (no bottom friction or refraction). The horizontal resolution of the wave model is 1.5º, angular resolution is 30º (12 directional bins), and the frequency resolution is logarithmic, spanning the range from 0.0420 to 0.4137 Hz. By keeping the model system and data assimilation system invariant over the 45-year integration period, the error statistics remain quite stationary, although the amount of observations is not constant over the period [*Uppala et al.*, 2005]. In particular the amount of satellite data has increased dramatically during the last two decades. The archive is relatively coarse compared with operational high-resolution data assimilation systems, but gives a good reproduction of most large-scale dynamical features [*Uppala et al.*, 2005].

We performed a regional downscaling of the ERA-40 reanalysis to produce detailed, atmospheric fields using a hydrostatic numerical weather prediction model, the High Resolution Limited Area Model (HIRLAM) version 6.4.2, see *Undén et al*. [2002] on 10-11 km resolution, hereafter known as HIRLAM10. The model domain is a rotated spherical grid with the South Pole positioned at 22º S, 040º W, see Figure 1. The domain is resolved by 248x400 grid points with 0.1º horizontal grid resolution. HIRLAM employs a semi-implicit semi-Lagrangian two-time level integration scheme for the integration of the model equations. The vertical was resolved by 40 hybrid levels with variable grid spacing (denser near the surface). The hybrid vertical co-ordinate transitions gradually from a pressure coordinate at the top to terrain-following coordinates near the surface. Further details on the discretization of the model equations, the handling of non-linear terms, diffusion and boundary relaxation can be found in *Undén et al*. [2002].

The model is forced by ERA-40 on the boundaries with temperature, wind, specific humidity and cloud water in all 40 model levels plus surface pressure with six-hourly temporal resolution. The fields are blended with the last forecast in the beginning of each cycle prior to the forecast integration as described below. Daily fields of sea surface temperature, sea ice fraction and snow depth are also retrieved from ERA-40. The ERA-40 sea surface temperature fields and sea ice fraction are extrapolated towards the higher resolution coasts in the HIRLAM10 grid by means of a grid-filling routine.

A sequence of 9-hourly model runs is performed starting from an initialization where ERA-40 is blended with the previous 6-h forecast field (the background field) in the interior of the domain. This is done to control the large-scale features of the forecasts [*Yang*, 2005]. The blending is generated by an incremental digital filter initialization (IDFI) scheme [*Lynch and Huang*, 1992 and *Lynch*, 1997] with a Dolph window filter. This digital filtering procedure is applied twice; once to the ERA-40 fields and once to the background fields. The initialization increment (the difference between the two filtered states) is then added to the background model state to obtain the initialized model state. The parameters adjusted by the incremental digital filter are pressure, wind, temperature, specific humidity and cloud water in all model levels. The incremental digital filtering initialization is meant to preserve quickly evolving modes in the first-guess [*Huang and Yang*, 2002].

## 2.2 The wave model configuration

The wave hindcast was generated using a nested setup of our modified version of the WAM Cycle 4 model [*WAMDI* 1988, *Komen et al.* 1994; *Günther et al.*, 1992] forced by HIRLAM10 winds on the same rotated spherical grid (hereafter known as WAM10). The model was nested inside a 50 km resolution WAM model (with identical numerics) covering most of the North Atlantic to ensure realistic swell intrusion from the North Atlantic (hereafter known as WAM50). This coarse model is forced with ERA-40 wind fields. The model domains are shown in Figure 1. We applied a new nesting procedure which allows arbitrary orientation of coarse and nested model domains [*Breivik et al.*, 2009]. The model is set up with twice as many directional bins as ERA-40 (15º, 24 bins) and 25 logarithmically spaced frequency bins from 0.0420 to 0.4137 . Ice coverage is updated three times per month from the HIRLAM10 ice cover. Both model integrations are run in shallow-water mode, *i.e.*, bottom friction is taken into account using realistic bottom topography.

## 3 Comparison of modeled wind and wave fields with observations and ERA-40

This study has focused on the quality of the wind and wave fields. The atmospheric downscaling has been compared to 10 m wind speed at coastal and offshore stations, and three years of QuikSCAT/SeaWinds scatterometer winds (years 1999-2002). The wave hindcast has been compared with *in situ* and satellite borne altimeter observations of significant wave height plus *in situ* mean wave period observations from wave buoys. We also compare ERA-40 against the same observations to assess to which extent the downscaling/hindcast represents an improvement.

## 3.1 Surface wind

A total of 22 coastal and oceanic weather stations with continuous and reliable observations over approximately the whole model period were used to evaluate the long-term performance of HIRLAM10 *v* ERA-40. In addition, several shorter records from offshore stations and some land stations were used for comparison of wind speed distributions, bringing the total up to 77 stations. Error: Reference source not found indicates that HIRLAM10 has a better overall performance in terms of 10 m wind speed. The mean absolute error (MAE) is 0.41 m s$^{-1}$ (17 % lower than ERA-40) and the root-mean-square error (RMSE) is 0.5 m s$^{-1}$ lower than ERA-40. ERA-40 is generally underestimating the 10 m wind speed with a mean error of -0.86 m s$^{-1}$, while HIRLAM10 has negligible bias. Of the 77 coastal, offshore and land stations, HIRLAM10 has a lower MAE compared to ERA-40 for 46 stations and exhibit a higher correlation with observed wind speed than ERA-40 at 57 stations. HIRLAM10 has a better maximum value than ERA-40 in 49 out of 77 stations.

As a brief assessment of the capability of the HIRLAM10 downscaling to capture polar lows we selected three representative polar lows where the maximum observed winds were in the range 23-26 m s$^{-1}$. The lows were identified manually by the Norwegian Meteorological Institute based on satellite imagery from the Advanced Very High Resolution Radiometer (AVHRR), synoptic observations and visual reports. Extremes often occur 3 to 9 h later than registered and the real mean sea level pressure (MSLP) is assumed to be within +/- 3 hPa of the recorded value. The results are summarized in Table 2. In neither case did HIRLAM10 capture the full evolution of the polar low, but wind speeds were higher than those from ERA-40 by up to 6 m s$^{-1}$ and a low was present for a while in all cases. We find in all three cases that HIRLAM10 manages to resolve the beginning of a polar low (with a diameter of 50-100 km), but because of the way the model is set up to go through short (9 h) cycles, the lows almost disappear upon reinitialization at the beginning of the next model cycle and do not develop into complete polar lows on a scale of 100-1000 km. ERA40 does not capture the low pressure centre at all. This is unsurprising as the smallest features represented in a spectral model have a horizontal scale dictated by the spectral truncation limit of the model.

### 3.1.1 Long term (45-year) average surface wind speed

Figure 2 shows the annual mean 10 m wind speed (vector-averaged) over the 45-year period in HIRLAM10 (Panel a) and the vector average of the difference between HIRLAM10 and ERA-40 (Panel b). For both HIRLAM10 and ERA-40 the westerlies south of Iceland away from continental influence are the strongest average winds in the model domain. HIRLAM10 generally exhibits much more fine-scale features in the coastal zone than the relatively smooth ERA-40 fields, owing to the more detailed topography in the model. The smallest mean differences (0.5 m s$^{-1}$ stronger winds in HIRLAM10) are found over the ocean while the largest differences are found at or near the continents (Panel b). The influence of the weaker winds in ERA-40 is seen along the open boundary in Panel (b). This is consistent with the bias found between ERA-40 and observations from both satellite-borne instruments (scatterometer) and synoptic observations, presented in the following sections.

Figure 3 shows the time series of monthly values of mean error (ME), mean absolute error (MAE) and root-mean-square error (RMSE) of 10 m wind speed at 22 coastal and offshore wind-measuring stations. HIRLAM10 shows substantially better performance compared with ERA-40 both in terms of bias (ME) and unsystematic deviations (MAE and RMSE). There was no trend in errors over the period. ERA-40 underestimated the 10 m wind speed (negative ME) while the RMSE and MAE were higher than for HIRLAM10. There is a clear annual cycle with the highest errors occurring in the winter in both datasets.

### 3.1.2 Wind speed distribution

The frequency distribution and associated quantile-quantile (qq) plot for HIRLAM10 and ERA-40 at Fruholmen lighthouse is shown in Figure 4 (Fruholmen is shown in Figure 5). HIRLAM10 has a wind speed distribution curve that compares well with observations, while ERA-40 slightly overestimates the lowest winds and underestimates the highest winds, as is also evident from inspection of the quantile distribution. HIRLAM10 also underestimates the highest wind speed values, but to a smaller extent. These results are comparable to the findings at other coastal stations. This particular station, Fruholmen lighthouse, was chosen to highlight the influence of the coastline on the wind speed distribution (wind enhancement) in HIRLAM10 compared with the much coarser topography in ERA-40. The improvement over ERA-40 is much more pronounced here compared with offshore stations (not shown).

### 3.1.3 Co-location of QuikSCAT/SeaWinds with HIRLAM10 and ERA-40

The SeaWinds scatterometer onboard the polar-orbiting QuikSCAT satellite is a real-aperture microwave radar. The empirical relationship used to infer 10 m wind vectors from the radar backscatter is based on a neutrally stratified atmosphere, which can usually be assumed over the ocean, except in situations of cold air outbreaks [*Portabella and Stoffelen*, 2001]. We have performed a co-location of the QuikSCAT/SeaWinds Level 2B Ocean Wind Vectors 25 km swath data obtained from the Physical Oceanography Distributed Active Archive Center (PO.DAAC) at the NASA Jet Propulsion Laboratory using a nearest-neighbor selection from the HIRLAM10 and the ERA-40 model grids to investigate the performance of the 10 m wind speed of the HIRLAM10 downscaling in open ocean conditions over the period June 1999 to August 2002. All measurements flagged with ice were removed, but no rain flag was used as the impact-based multidimensional histogram (IMUDH) is found to over-predict the rain contamination, especially at high wind speed [*Portabella and Stoffelen,* 2002]. We found a significant improvement in bias from the ERA-40 mean error of -0.77 m s$^{-1}$ to -0.02 m s$^{-1}$ for HIRLAM10. Similarly, the RMS error went down from 2.32 m s$^{-1}$ for ERA-40 to 2.08 m s$^{-1}$ for HIRLAM10 (see summary of wind statistics in Error: Reference source not found). It is known that QuikSCAT wind measurements are biased high for very high wind speeds [above 19 m s$^{-1}$, see *Ebuchi et al.*, 2002; *Moore et al.*, 2008]. When used in assimilation the QuikSCAT winds are reduced overall by 4% while winds above 19 m s$^{-1}$ are reduced according to the formula $V'=v-0.2(v-19)$, where $v$ is the uncorrected wind speed and $V'$ is the corrected wind speed [*ECMWF*, 2002]. Figure 6 compares QuikSCAT and HIRLAM10 wind speed over the range 0-20 m s$^{-1}$, where QuikSCAT can be used uncorrected. Strong correlation ($r=0.88$) is found between the downscaled wind fields and QuikSCAT with the bulk of the observations exhibiting near 1:1 correspondence (very low bias). Over this wind-speed range the quantiles appear relatively similar although we note a slight deviation in the quantiles above 15 m s$^{-1}$. Figure 6 also includes the qq plot of QuikSCAT $v$ ERA-40. Although ERA-40 wind fields also correlate well ($r=0.86$) there is an underestimation of about 10% for winds higher than 3 m s$^{-1}$ (see also Error: Reference source not found). To further investigate the geographical differences of the downscaling we looked at the winter months December, January and February (DJF), when the differences are largest. The RMSE between QuikSCAT and HIRLAM10 shown in Figure 7 reveals that the largest deviations occur south and east of Svalbard in the winter months, most probably in connection with the ice edge. The elevated RMSE in this region may result from a combination of few observations, poor resolution of the true ice edge in the model configuration and ice in the field of view of the scatterometer (erroneous observations may occasionally slip through the ice flag algorithm), *i.e.*, errors are expected to be higher both in the modeled fields and the satellite observations in the marginal ice zone. The mean difference is shown in Figure 8. There is generally good agreement between the HIRLAM10 downscaled wind fields and the satellite measurements, but the influence from the ERA-40 (biased low) manifests itself in a zone of weaker winds near the open boundaries (consistent with the differences found in Figure 2b). There are four regions where HIRLAM10 predicts higher wind in the mean than observed by QuikSCAT: the Southern North Sea, North-East of Iceland, in the Greenland Sea and near Bjørnøya.

## 3.2 Wave field

### 3.2.1 Significant wave height

Two *in-situ* datasets were used for the assessment of the significant wave height, one consisting of six offshore stations in the Norwegian sector quality-checked and archived by The Norwegian Meteorological Institute and another dataset comprising 40 wave observing buoys, coastal and offshore stations (Figure 5) collected via the Global Telecommunications System (GTS) and quality-checked and archived by ECMWF [*Saetra and Bidlot* 2004]. We treat the two datasets separately as they have been subjected to different methods of quality assurance and temporal averaging. In addition, a two-year co-location of satellite-borne altimeter measurements of significant wave height against ERA-40 and WAM10 has been carried out.

Table 3 and Table 4 compare the observed and modeled significant wave height at the six Norwegian offshore stations from Ekofisk at 56.5º N, 003.2º E in the central North Sea to Ami at 71.5º N, 019.0º E off the coast of Northern Norway over the whole observational period. Of these, Ekofisk has the longest and most reliable time series, starting in 1980. The WAM10 wave fields correlate very well with Ekofisk observations (0.95), show a low RMS error (0.42 m) and no bias, see Table 3. In general WAM10 appears to overestimate the 99 percentile (P99) by 0.1 to 0.6 m (average of 0.3 m). ERA-40 wave fields (Table 4) also match the observed time series well in all six stations, with correlations only slightly weaker than WAM10. However, the mean wave height is consistently underestimated, ranging from 10 to 34 cm. Furthermore, the RMS error is higher than for WAM10. The disagreement with observed values becomes increasingly worse for the higher percentiles. ERA-40 is underestimating P99 by 0.6 to 1.3 m (Draugen).

We have also compared WAM10 and ERA-40 against a collection of 40 in-situ wave observing stations from the North-East Atlantic Ocean (see Figure 9 and Table 5) during the period August 1991 – August 2002. The individual time series vary in length from half a year to slightly more than nine years. All observations are quality-controlled four-hourly means (±2 h) centered on synoptic times, i.e. 00, 06, 12 and 18 UTC [*Saetra and Bidlot,* 2004]. Most measurements are permanently fixed to one location, but some buoys exhibit rather large deviations from their average position. Here we used only data within an area of ± 0.2° of the median latitude and ± 0.4° of the median longitude of the individual time series. Figure 9 compares observed and modeled significant wave height ($H_s$) from WAM10 and ERA-40, showing a one to one correspondence between WAM10 and observations. ERA-40 overestimates the lower percentiles while underestimating the upper percentiles, yielding a rather weak regression slope of 0.84. This deficiency is masked when looking at the mean error (bias -0.01 m *v* 0.08 m for WAM10), but the RMS error is substantially higher than for WAM10 (0.55 v 0.47). The scatter index (SI), defined as the RMS error normalized by the mean of the observations drops from 24.7% for ERA-40 to 20.3% for WAM10. Figure 10 compares observed and modeled 95 and 99-percentiles from ERA-40 and WAM10 at all 40 observing stations. As **Figure 1** testifies, the wave climate differs significantly between these observing stations with the January mean ranging from 1.5 m for stations close to the Dutch coast to more than 4.5 m south of Iceland. It is therefore relevant to investigate how the upper percentiles of the WAM10 hindcast compare with observations from such a diverse range of locations. It is evident from Figure 10 that the upper percentiles of WAM10 match observations better than ERA-40 in nearly all 40 observing stations with better correlation for P99 (0.95 *v.* 0.88) and a regression slope closer to unity. It appears that low-wave locations are overestimated and high-wave locations are underestimated by ERA-40, somewhat analogously to the overall pattern found in Figure 9. This is consistent with the findings for the Norwegian offshore stations (Table 3 and Table 4). Overall, WAM10 is again found to be biased a little high for the upper percentiles (a normalized mean error of 5 % for P95), while ERA-40 is biased a little low (-5 % for P95), as summarized in Table 5.

WAM10 was also co-located with significant wave height from altimeters onboard ERS-2, Topex and Geosat Follow-On (GFO) for the years 2000-2001. The temporal resolution of the gridded fields is 3 hours, hence the maximum time difference between satellite observations and modeled significant wave height is 1.5 hours and the maximum spatial separation is approximately 7 km. As can be seen from Figure 11, the overall agreement is very good with a correlation of 0.95 (Table 6) and very low bias. A similar co-location was performed for ERA-40 (the lower quantile curve, shown in Figure 11 for comparison with the quantiles of WAM10, upper curve). ERA-40 correlates strongly with altimeter wave height (0.94), but underestimates the highest waves somewhat as is evident from the qq-plot. Figure 11 shows that while the quantiles of WAM10 (black line) are almost unbiased up to 9 m, ERA-40 underestimates waves higher than 2.5 m while overestimating waves below 2.5 m. This is consistent with the comparison against *in situ* observations (Table 3-Table 5). Note that while WAM10 slightly overestimates the upper percentiles compared with buoy observations (Figure 9), an underestimation is found when comparing with altimeter data (Figure 11).

### 3.2.2 Mean wave period

A comparison of the measured mean period (both zero upcrossing, $T_z$, and $T_{m02}$ based on observed wave spectra are used) and the modeled $T_{m02}$ calculated from the two-dimensional spectrum was carried out for a subset of the stations listed in Figure 5 where reliable wave period estimates could be made. The greatest deviations between modeled and observed periods appear in the upper tail (Figure 12), but ERA-40 also exhibits large deviations in cases with relatively short observed wave periods. This may indicate that there are certain swell cases that are not well captured by the models, but also that low-wind cases are clearly better represented by WAM10. Overall the mean period was rather well represented by WAM10 with a correlation of 0.92. ERA-40 showed a weaker correlation (0.84).

# 4 Discussion

The downscaling of ERA-40 using HIRLAM10 with 10-11 km horizontal resolution shows a significant improvement in 10 m wind. Although the improvement is particularly pronounced at coastal stations, we also found that the negative wind speed bias of ERA-40 disappeared throughout the open ocean in the HIRLAM10 downscaling. Improvement through dynamical downscaling in coastal areas and areas with complex topography has been shown in many studies before; for example by *Winterfeldt* [2008] and *Winterfeldt et al.* [2010]. However, the improvement found at offshore stations and from the QuikSCAT co-location in the open ocean is not usually expected in a dynamical downscaling, although similar results have been found by *Feser* [2006]. Some of the regional differences between HIRLAM10 and QuikSCAT found in Figure 7 and Figure 8, such as the wind minimum in the Greenland Sea, was also noted by *Kolstad* [2008] when comparing QuikSCAT to NCEP/NCAR reanalysis data. The differences are in part due to the assumption of neutral stratification when estimating the scatterometer wind speed from regions with non-neutral near-surface stratification [*Chelton and Freilich*, 2005; *Accadia et al.*, 2007; *Furevik et al.,* 2010].

Though we can identify polar lows in HIRLAM10 which are unresolved by ERA-40, the downscaling is not modeling the full evolution of such small-scale cyclones. Due to the nature of the HIRLAM10 downscaling, performed as a sequence of short (six-hour) forecasts strongly influenced by the ERA-40 boundary conditions and the digital filter initialization procedure, a good representation of polar lows is not likely. Improved modeling of polar lows would probably require even higher horizontal and vertical resolution and non-hydrostatic modeling. A more sophisticated spatial filter such as the spectral nudging technique employed by *Zahn et al.* [2008] in a limited area model of 0.44º resolution might be capable of capturing and resolving polar lows not present in the coarse boundary conditions and initial conditions. Scale-selective bias correction as explored by *Kanamaru and Kanamitsu* [2006] in a regional downscaling of NRA is another promising technique, but as no observations are assimilated into our dynamical downscaling it is difficult to assess whether this technique would have fared better than our method with frequent reinitializations and the use of a digital filter. It is not expected that estimates of wind speed extremes will be seriously affected by this inability to correctly capture and model polar lows as their associated wind speed will generally not be in the upper tail of the wind speed distribution in the Barents Sea and the northern Norwegian Sea. Around Svalbard the performance of both datasets is lower than in the rest of the area. HIRLAM10 shows however somewhat lower winter-time MAE and RMSE than ERA-40, but both HIRLAM10 and ERA-40 overestimate 10 m wind speed in this area without properly capturing the extreme events. This is probably due to imprecise ice cover representation.

The significant wave height of the WAM10 hindcast has a smaller mean error than ERA-40 which was found to be biased low both for the mean and for the upper percentiles (though partly compensated by a positive bias for the lower wave heights, see Figure 9). Overall, ERA-40 seems to underestimate the upper percentiles of the significant wave height by approximately 10 % while WAM10 overestimates the highest percentiles by about 6 % in comparison with buoys and offshore stations (Figure 10, Table 3 and Table 4). However, the comparison with altimeter wave height reveals that WAM10 remains unbiased up to 9 m while ERA-40 underestimates waves higher than 2.5 m by about 10 % (Figure 11). These findings are broadly similar to what *Sterl and Caires* [2005] found for ERA-40. WAM10 shows significant improvement over ERA-40 in terms of both correlation and regression slope. There is also a clear improvement in the modeled mean period compared with ERA-40 (Figure 12b). The results are similar to the findings of *Caires et al.* [2005] who performed a global comparison of ERA-40 $T_m$ with estimates based on TOPEX altimeter observations and buoy data. Due to limitations in the altimeter measurements their main focus was on non-swell-dominated cases, *i.e.* $H_{Swell}/H_S < 0.9$, but the results still appear comparable. It is interesting to note that WAM10 outperforms ERA-40 even for the long-periodic observations in the North Atlantic where it might be expected that intrusion of swell from the South Atlantic could influence the results (the WAM50 model domain extends only into the North Atlantic, see Figure 1).

As WAM10 was nested in a coarser model, WAM50, that differed somewhat from the ERA-40 wave model, we investigated the possible impact this has had by comparing WAM50 and ERA-40 in terms of significant wave height in a representative range of buoy locations in the North Atlantic and the North Sea (not shown). We found very close correspondence between WAM50 and ERA-40 at buoy locations 62108, 64045 in the North Atlantic and at Ekofisk in the central North Sea (see Figure 5) and infer that the improvement found in WAM10 stems primarily from better wind fields and a more detailed topography. This also means that doubling the directional resolution (WAM50 has 24 directional bins) and employing bottom friction (ignored by ERA-40) makes little impact on wave model integrations on the spatial resolution of ERA-40. However, both directional resolution and bottom friction become important near shore in a high-resolution model domain. WAM10 outperforms ERA-40 throughout the model domain, but although we have few wave observations in the coastal zone and in shallow areas, we expect the improvement to be higher here due to the demonstrably better near-shore wind fields and the improved coastline representation.

## 5 Conclusion

This study has shown that major improvement of coastal and open-ocean wind fields can be achieved through a relatively straight forward dynamical downscaling of the ERA-40 reanalysis. The method does not rely on additional assimilation but still reduces the mean error and the RMS error at coastal stations dramatically (up to 3 m s$^{-1}$ reduction in mean error near Stad on the coast of Norway). The improvement near the coast was expected from the enhanced resolution alone, but the improvement in the open ocean was more surprising as there is no additional source of observations to correct the downscaled wind fields. The overall impression of the open ocean wind field of the HIRLAM10 downscaling is a marked improvement compared with ERA-40. This can only be explained by better mesoscale representation of the weather systems in HIRLAM10. Further work is still needed to properly address the downscaling of mesoscale phenomena such as polar lows, and for this a scale-selective approach to downscaling may be required. However, the main concern from the point of view of dynamically downscaling the ERA-40 reanalysis is to reconstruct the wind field as truthfully as possible without generating small-scale non-existent disturbances. This is a difficult trade-off, but one where a digital filter does a good job, albeit at the expense of modeling the full evolution of polar lows.

The wave model integration has benefited from the improved wind fields both in terms of significant wave height, but also in the representation of the mean wave period. The study shows that although the coarse model does not cover the South Atlantic, WAM10 exhibits more correct wave periods even in the exposed parts of the northern North Atlantic when compared against ERA-40. As for the wind fields, marked improvement is found both in the distribution of the upper percentiles of the significant wave height as well as the overall frequency distribution.

This study has yielded a new high-resolution wave and wind archive of markedly improved quality through a relatively straight forward dynamical downscaling of the ERA-40 reanalysis for an important region of the world oceans. The wind and wave fields are virtually un-biased in coastal regions where ERA-40 alone is not usable and also improved in the open ocean. Although some of the shortcomings of ERA-40 have already been addressed by the ERA Interim reanalysis from 1989 to present [*Simmons et al*, 2010], the period covered is still too short to replace the ERA-40 reanalysis, and the resolution is still too coarse to adequately resolve coastal features and mesoscale phenomena such as polar lows. Although we emphasize the need for better global reanalyses, we conclude that ERA-40 represents a good starting point for dynamical regional downscaling of wind and wave fields which yields genuine new information about near-shore features as well as somewhat improved mesoscale activity and a better representation of mean conditions, even in the open ocean. The significant improvement found in the upper tail of the wind and wave height distribution also makes the downscaled fields more reliable for extreme value statistics than the original ERA-40 reanalysis.

## Acknowledgments

This project was funded by the Norwegian Deepwater Programme (www.ndwp.org). The project has relied on computing resources of the Norwegian metacenter for computational science, NOTUR. The QuikSCAT/SeaWinds Level 2B Ocean Wind Vectors 25 km swath data was obtained from PO.DAAC at the NASA Jet Propulsion Laboratory, Pasadena, CA, http://podaac.jpl.nasa.gov. The GFO, TOPEX and ERS-2 altimeter data were obtained from CERSAT, IFREMER, France.

# References

Accadia, C., S. Zecchetto, A. Lavagnini and A. Speranza (2007), Comparison of 10 m wind forecasts from a regional area model and QuikSCAT scatterometer wind observations over the Mediterranean Sea. *Mon. Wea. Rev, 135*, 1945-1960

Breivik, Ø, Y Gusdal, B R Furevik, O J Aarnes and M Reistad (2009), Nearshore wave forecasting and hindcasting by dynamical and statistical downscaling, *J Marine Sys*, *78*, 235-243, doi:10.1016/j.jmarsys.2009.01.025

Bromwich, D. H., R. L. Fogt, K .I. Hodges and J. E. Walsh (2007), A tropospheric assessment of the ERA-40, NCEP and JRA-25 global reanalyses in the polar regions, *J. Geophys. Res*, *112*, D10111, doi:10.1029/2006JD007859

Caires S and A Sterl (2005), 100-year return value estimates for ocean wind speed and significant wave height from the ERA-40 data. *J. Climate 18*, 1032–1048

Caires, S., A. Sterl and C.P. Gommenginger (2005), Global ocean mean wave period data: Validation and description. *J. Geophys. Res. 110*, C02003, doi:10.1029/2004JC002631

Chelton, D. B., and M. H. Freilich (2005), Scatterometer-based assessment of 10 m wind analysis from the operational ECMWF and NCEP numerical weather prediction models, *Mon. Wea. Rev.*, *133*, 409-429

Condron A., G.R. Bigg and I. A. Renfrew (2005), Polar Mesoscale Cyclones in the Northeast Atlantic: Comparing Climatologies from ERA-40 and Satellite Imagery, *Mon. Wea. Rev.*, *134*, 1518-1533.

Cox, AT and Greenwood, JA and Cardone, VJ and Swail, VR (1995), An interactive objective kinematic analysis, *Proc. Fourth Int. Workshop on Wave Hindcasting and Forecasting*, 109-118

Cox, A. T., and V. R. Swail (2001), A global wave hindcast over the period 1958–1997: Validation and climate assessment, *J. Geophys. Res.*, *106*(C2), 2313–2329

Ebuchi, N., H.C. Graber, and M.J. Caruso (2002), Evaluation of Wind Vectors Observed by QuikSCAT/SeaWinds Using Ocean Buoy Data. *J. Atmos. Oceanic Technol.*, *19*, 2049–2062.

ECMWF (2002), *IFS Documentation, CY25R1 Operational implementation 9 April 2002*, Available online at http://www.ecmwf.int/research/ifsdocs/CY25r1/index.html or http://tinyurl.com/ygb7fu4 (Accessed 10 November 2009).

Eide, L.I, M.Reistad, J.Guddal (1985), *A database of computed wind and wave parameters for the North Sea , The Norwegian Sea and the Barents Sea for every 6 hours for the years 1955-1981* (In Norwegian). Norwegian Meteorological Institute, project report.

Feser. F. (2006), Enhanced detectability of added value in limited-area model results separated into different spatial scales. *Mon. Wea. Rev.*, *134*, 2180-2190.

Furevik, B. R., A. M. Sempreviva, L. Cavaleri, J.-M. Lefevre and C. Transerici (2010), Eight years of wind measurements from scatterometer for wind resource mapping in the Mediterranean Sea, *Wind Energy*, DOI: 10.1002/we.425

Garcia-Sotillo, M., A. Ratsimandresy, J. Carretero, A. Bentamy, F. Valero, and F. Gonzalez-Rouco (2005), A high-resolution 44-year atmospheric hindcast for the Mediterranean Basin: Contribution to the regional improvement of global reanalysis. *Climate Dyn.*, *25*(2-3), 219-236, doi:10.1007/s00382-005-0030-7

Gaslikova L and Weisse R (2006), Estimating near-shore wave statistics from regional hindcasts using downscaling techniques. *Ocean Dynamics 56*, 26–35

Gibson, JK, P. Kållberg and S. Uppala (1996), The ECMWF re-analysis (ERA) project, *ECMWF newsletter*, *73*, 7-17

Gibson, JK, P. Kållberg, S. Uppala, A. Hernandez, A. Nomura and E. Serrano (1997), *ERA-15 description*, ECMWF Reanalysis Project Report Series, *1*, 72 pp

Günther, H., S. Hasselmann and P.A.E.M. Janssen (1992), *The WAM model cycle 4*. Technical Report, Deutsches KlimaRechenZentrum, Hamburg, Germany

Günther, H., Rosenthal, W., Stawarz, M., Carretero, JC, Gomez, M., Lozano, I., Serrano, O. and Reistad, M (1998), The wave climate of the Northeast Atlantic over the period 1955-1994: The WASA wave hindcast, *The Global Atmosphere and Ocean System*, *6*(2), 121-163

Haug, O. and J. Guddal (1981), *Hindcasting the Wind and Wave Climate of Norwegian Waters*. Norwegian Meteorological Institute, project report.

Huang X-Y, and X. Yang (2002), *A new implementation of digital filtering initialization schemes for HIRLAM*. DMI Technical Report.

Kalnay, E., M Kanamitsu, R Kistler, W Collins, D Deaven, L Gandin, M Iredell, S Saha, G White, J Woollen *et al.* (1996), The NCEP/NCAR 40-year reanalysis project, *Bulletin of the American Meteorological Society*, *77*(3), 437-471

Kanamaru, H. and M. Kanamitsu (2006), Scale-Selective Bias Correction in a Downscaling of Global Analysis Using a Regional Model, *Mon. Wea. Rev.*, *135*, 334-350.

Kolstad, E. (2008), A QuikSCAT climatology of ocean surface winds in the Nordic Seas: Identification of features and comparison with the NCEP/NCAR reanalysis, *J. Geophys Res.*, *113*, D11106, DOI:10.1029/2007JD008918

Komen, G.J., L. Cavaleri, M. Donelan, K. Hasselmann, S. Hasselmann, and P.A.E.M. Janssen (1994), *Dynamics and Modelling of Ocean Waves*. Cambridge University Press, Cambridge.

Laprise, R. (1992), The resolution of global spectral models, *Bull. Amer. Meteor. Soc.*, *73*, 1453-1454.

Lopatoukhin, LJ, VA Rozhkov, VE Ryabinin, VR Swail, AV Boukhanovsky and AB Degtyarev (2000), *Estimation of extreme wind wave heights*, JCOMM Technical Report No 9, World Meteorological Organization

Lynch, P and X.-Y. Huang (1992), Initialization of the HIRLAM model using a digital filter. *Mon. Wea. Rev., 120*, 1019–1034

Lynch, P. (1997), The Dolph-Chebyshev Window: A Simple Optimal Filter, *Mon. Wea. Rev., 125*, 655-660.

Moore, GW, KRS Pickart, IA Renfrew (2008), Buoy observations from the windiest place in the world ocean, Cape Farewell, Greenland, *Geophys. Res. Lett., 35*, doi: 10.1029/2008GL034845.

Onogi, K., H Koide, M Sakamoto, S Kobayashi, J Tsutsui, H Hatsushika, T Matsumoto, N Yamazaki, H Kamahori, K Takahashi, *et al.* (2005), JRA-25: Japanese 25-year re-analysis project-progress and status, *Q. J. Roy. Meteorol. Soc., 131*(613), 3259-3268.

Pielke, R.A. (1991), Recommended specific definition of "resolution". *Bull. Amer. Meteor. Soc., 72*, 1914

Pilar, P., C. Guedes Soares and J.C. Carretero (2008), 44-year wave hindcast for the North East Atlantic European coast, *Coastal Engineering, 55* (11), 861-871, DOI: 10.1016/j.coastaleng.2008.02.027

Portabella, M., and A. Stoffelen (2001), Rain Detection and Quality Control of SeaWinds. *J. Atmos. Oceanic Technol., 18*, 1171–1183

Portabella, M. and A. Stoffelen (2002), A comparison of KNMI quality control and JPL rain flag for SeaWinds. *Can J Rem Sens, 28*, 424–430.

Ratsimandresy, A.W., M.G. Sotillo, J.C. Carretero Albiach, E. Alvarez Fanjul, H. Hajji (2008), A 44-year high-resolution ocean and atmospheric hindcast for the Mediterranean Basin developed within the HIPOCAS Project, *Coastal Eng, 55*(11), DOI: 10.1016/j.coastaleng.2008.02.025

Reistad, M. and K. Iden (1998), *Updating, correction and evaluation of a hindcast data base of air pressure, wind and waves for the North Sea, the Norwegian Sea and the Barents Sea*. Norwegian Meteorological Institute Research report No. 9

Reistad, M, Ø Breivik and H Haakenstad (2007), A High-Resolution Hindcast Study for the North Sea, the Norwegian Sea and the Barents Sea, *in Proceedings of the 10th International Workshop on Wave Hindcasting and Forecasting and Coastal Hazard Symposium*, 13pp

Saetra, O. and Bidlot, J. R. (2004), On the potential benefit of using probabilistic forecast for waves and marine winds based on the ECMWF ensemble prediction system. *Wea. Forecasting 19*, 673-689.

Simmons, A. J., K. M. Willett, P. D. Jones, P. W. Thorne, and D. P. Dee (2010), Low-frequency variations in surface atmospheric humidity, temperature, and precipitation:

Inferences from reanalyses and monthly gridded observational data sets, *J. Geophys. Res., 115*, D01110, doi:10.1029/2009JD012442

Sterl, A., G. J. Komen, and P. D. Cotton (1998), Fifteen years of global wave hindcasts using winds from the European Centre for Medium-Range Weather Forecasts reanalysis: Validating the reanalyzed winds and assessing the wave climate, *J. Geophys. Res. 103*(C3), 5477–5492.

Sterl, A and S Caires (2005), Climatology, variability and extrema of ocean waves: the Web-based KNMI/ERA-40 wave atlas, *Int. J. of Climatol. 25*(7), 963-977

Swail, V.R., and A.T. Cox (2000), On the Use of NCEP–NCAR Reanalysis Surface Marine Wind Fields for a Long-Term North Atlantic Wave Hindcast. *J. Atmos. Oceanic Technol., 17*, 532–545.

Undén, P., Rontu, L., Järvinen, H., Lynch, P., Calvo, J., Cats, G., Cuaxart, J., Eerola, K., Fortelius, C., Garcia-Moya, J.A., Jones, C., Lenderlink, G., McDonald, A., McGrath, R., Navascues, B., Nielsen, N.W., Ødegaard, V., Rodriguez,E., Rummukainen, M., Rööm, R., Sattler, K., Sass, B.H., Savijärvi, H., Schreur, B.W., Sigg, R., The, H. and Tijm, A. (2002), *HIRLAM-5 Scientific Documentation, HIRLAM-5 Project.* Available from SMHI, S-601767 Norrköping, Sweden.

Uppala, S.M., Kållberg, P.W., Simmons, A.J., Andrae, U., da Costa Bechtold, V., Fiorino, M., Gibson, J.K., Haseler, J., Hernandez, A., Kelly, G.A., Li, X., Onogi, K., Saarinen, S., Sokka, N., Allan, R.P., Andersson, E., Arpe, K., Balmaseda, M.A., Beljaars, A.C.M., van de Berg, L., Bidlot, J., Bormann, N., Caires, S., Chevallier, F., Dethof, A., Dragosavac, M., Fisher, M., Fuentes, M., Hagemann, S., Hólm, E., Hoskins, B.J., Isaksen, L., Janssen, P.A.E.M., Jenne, R., McNally, A.P., Mahfouf, J.-F., Morcrette, J.-J., Rayner, N.A., Saunders, R.W., Simon, P., Sterl, A., Trenberth, K.E., Untch, A., Vasiljevic, D., Viterbo, P., and Woollen, J. (2005), The ERA-40 re-analysis. *Quart. J. R. Meteorol. Soc., 131*, 2961-3012.

WAMDI group - Hasselmann, S, K. Hasselmann, E. Bauer, L. Bertotti, C. V. Cardone, J. A. Ewing, J. A. Greenwood, A. Guillaume, P. A. E. M. Janssen, G. J. Komen, P. Lionello, M. Reistad, and L. Zambresky (1988), The WAM Model - a third generation ocean wave prediction model, *J. Phys. Oceanogr., 18*(12), 1775-1810.

Wang, X.L., and V.R. Swail (2001), Changes of Extreme Wave Heights in Northern Hemisphere Oceans and Related Atmospheric Circulation Regimes. *J. Climate, 14*, 2204–2221

Wang, X.L., and V.R. Swail (2002), Trends of Atlantic Wave Extremes as Simulated in a 40-Yr Wave Hindcast Using Kinematically Reanalyzed Wind Fields. *J. Climate, 15*, 1020–1035

WASA group (1998), Changing waves and storms in the Northeast Atlantic? *Bull Am Meteorol Soc 79*, 741–760

Weisse R, Feser F, Günther H (2002), A 40-year high-resolution wind and wave hindcast for the Southern North Sea. *In: Proceedings of the 7th international workshop on wave hindcasting and forecasting*, Banff, Alberta, Canada, 21–25 October 2002, pp 97–104

Weisse, R and H Günther (2007), Wave climate and long-term changes for the Southern North Sea obtained from a high-resolution hindcast 1958–2002, *Ocean Dynamics*, *57*(3), 162-171

Weisse, R. and H von Storch (2010), *Marine Climate and Climate Change: Storms, Wind Waves and Storm Surges*, 200 p, Springer Praxis Books

Winterfeldt, J. (2008), *Comparison of Measured and Simulated Wind Speed Data in the North Atlantic*, GKSS Reprot 2008/2. GKSS Forschungszentrum, Geesthacht, Germany.

Winterfeldt, J., B. Geyer and R. Weisse (2010), Using QuikSCAT in the added value assessment of dynamically downscaled wind speed, *Int. J. Climatol*. doi: 10.1002/joc.2105

Yang, X. (2005), Background blending using an incremental spatial filter, *Hirlam Newsletter*, *49*, 3-11

Zahn, M., H von Storch, and S Bakan (2008), Climate mode simulation of North Atlantic polar lows in a limited area model, *Tellus A*, *60*(4), 620-631

**Figures**

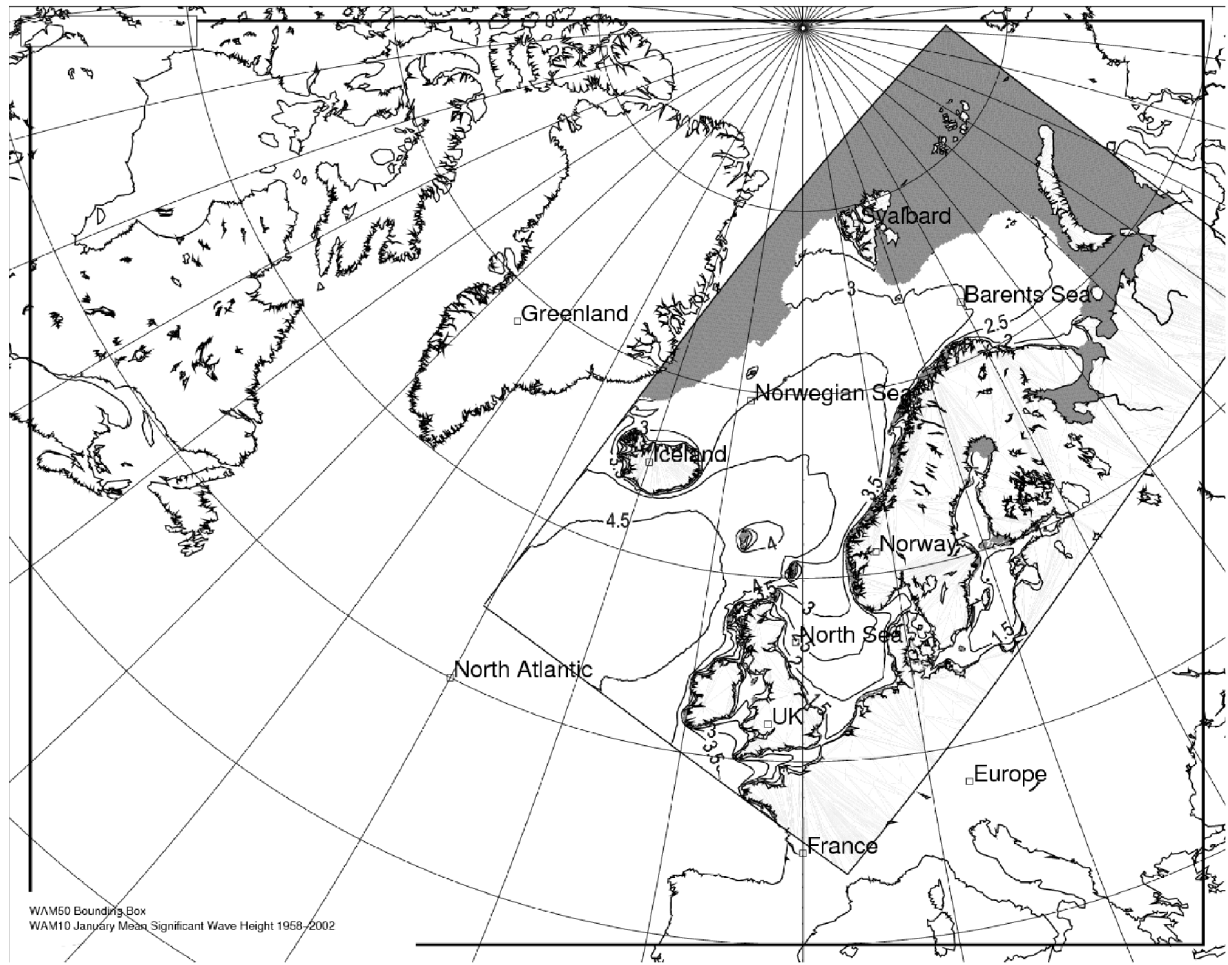

Figure 1. Model domains (10º graticule). The long-term WAM10 mean significant wave height [m] for January, 1958-2002, is shown with median ice extent in gray. The HIRLAM10 and WAM10 model domains are identical. The outer box indicates the WAM50 model domain. This domain is found to be sufficiently large to account for most swell intrusions into the WAM10 domain. Both domains are rotated spherical projections. The model projection of HIRLAM10/WAM10 has the south pole positioned at 22ºS, 040ºW with a 248x400 grid on 10-11 km resolution.

(a)

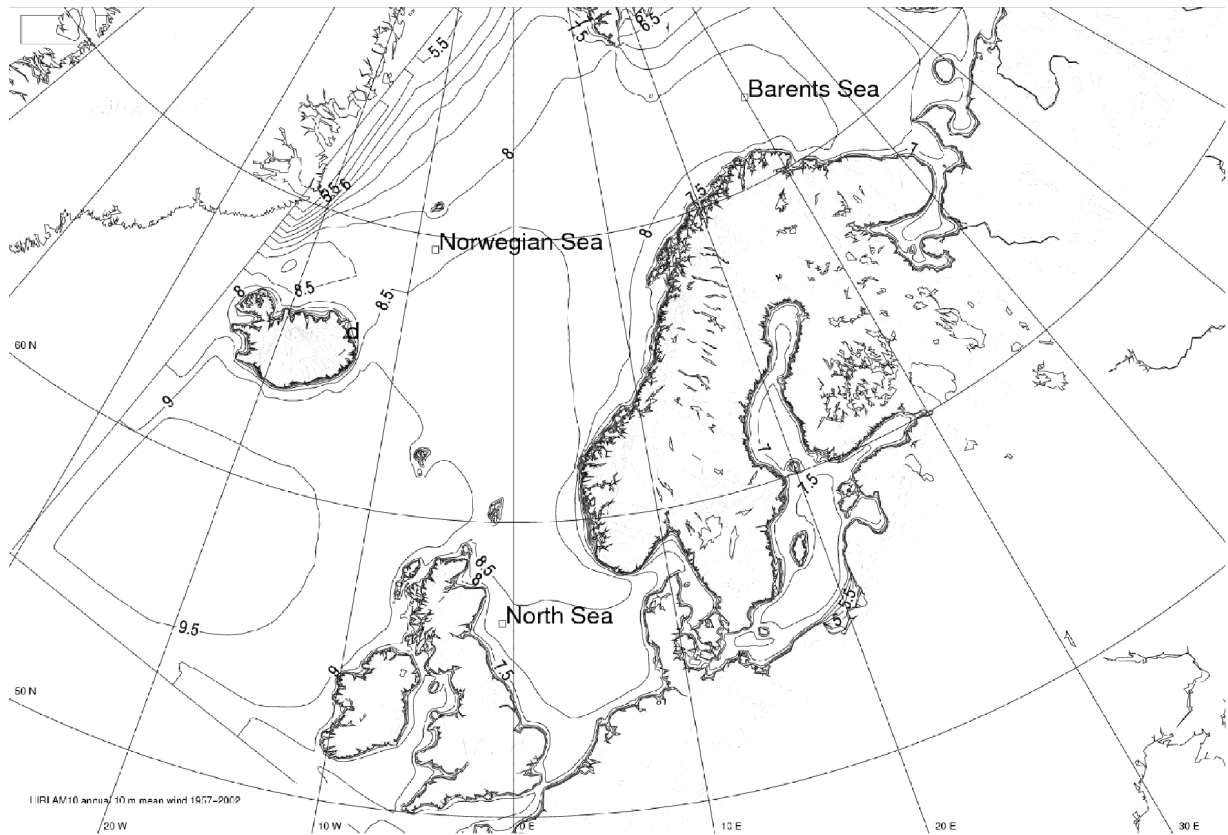

(b)

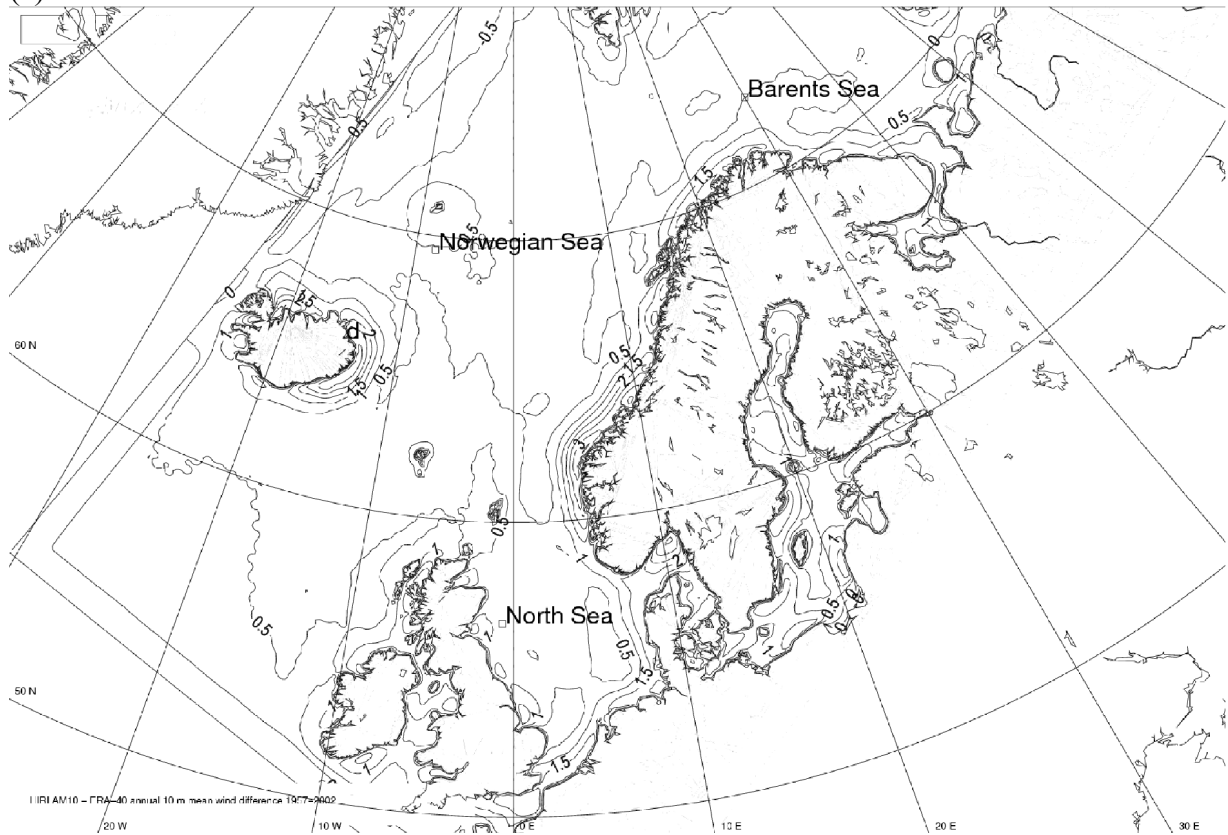

Figure 2. Mean 10 m wind vector-averaged over the period Sep 1957 to Aug 2002 for HIRLAM10 (Panel a) and their mean vector difference (HIRLAM10-ERA40, Panel b).

HIRLAM10 is consistently 0.5 m s$^{-1}$ higher in the open ocean except along the open boundary where the model is heavily influenced by ERA-40. The difference increases towards the coasts with a maximum of 3 m s$^{-1}$ near Stad at 62.5ºN on the coast of Norway, a region known for its funneling effect due to the sharp bend of the coastline.

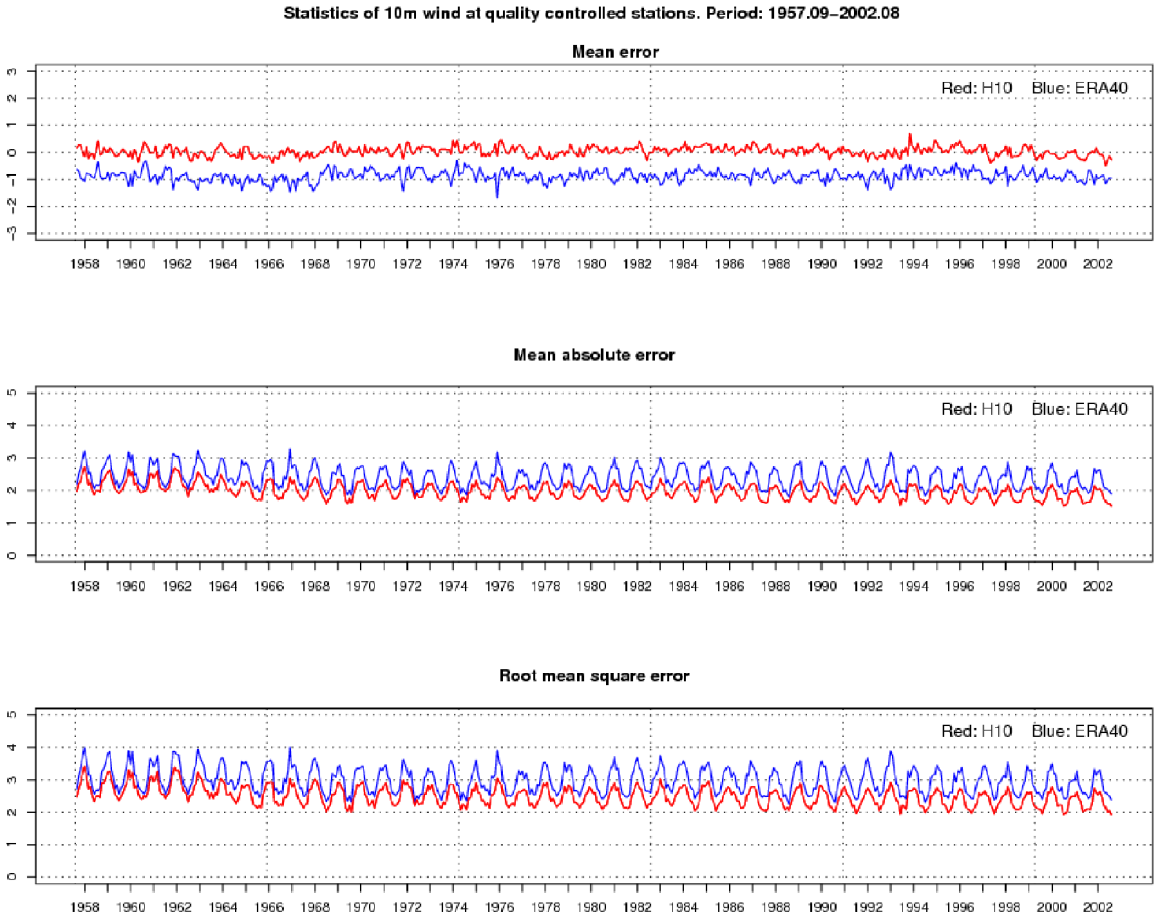

Figure 3. Time series (Sep 1957 to Aug 2002) of mean error, mean absolute error and root mean square error of 10 m wind speed. HIRLAM10 in red, ERA-40 blue.

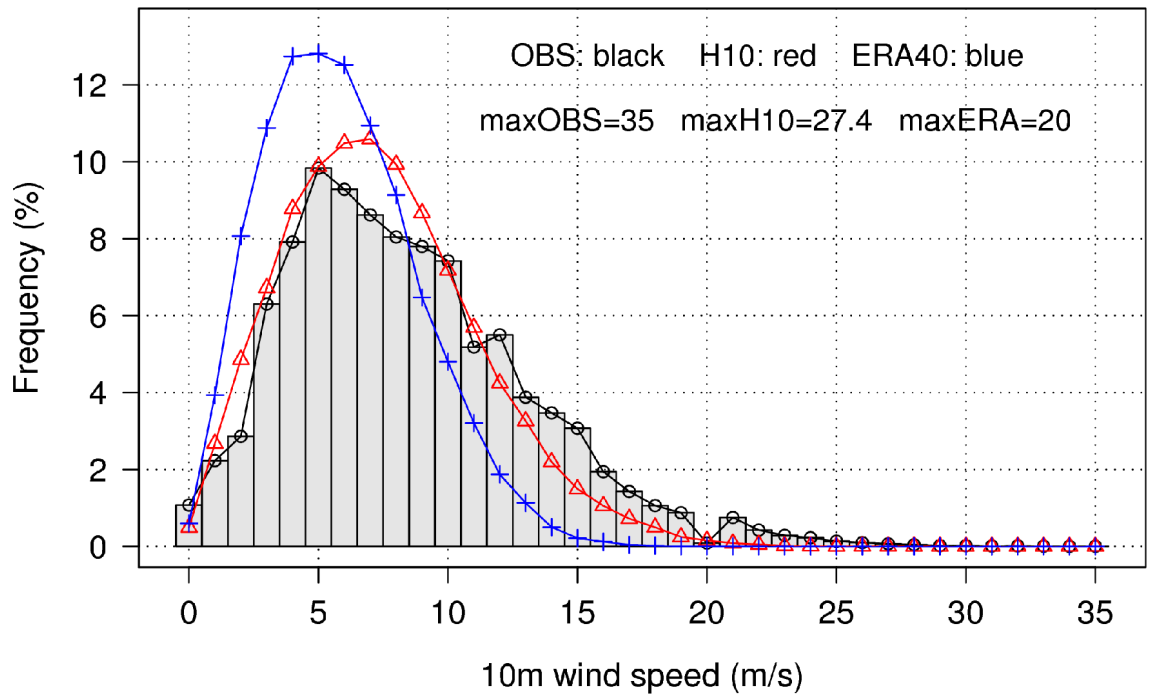

(a)

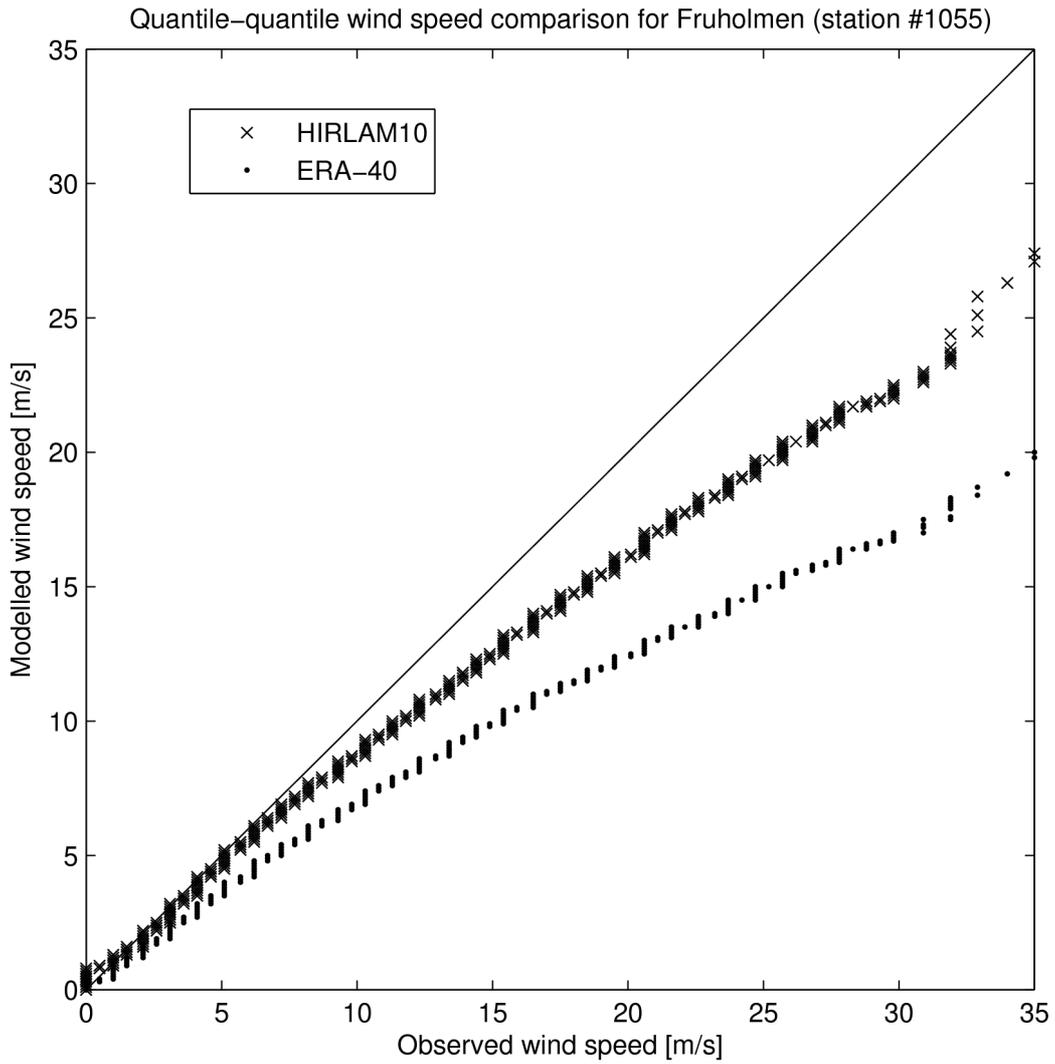

(b)

Figure 4. Upper panel (a): The distribution plot of 10 m wind speed for Fruholmen light house (HIRLAM10 in red, ERA-40 blue and observations as histogram). Lower panel (b): HIRLAM10 (crosses) and ERA-40 (dotted) quantiles v observed quantiles.

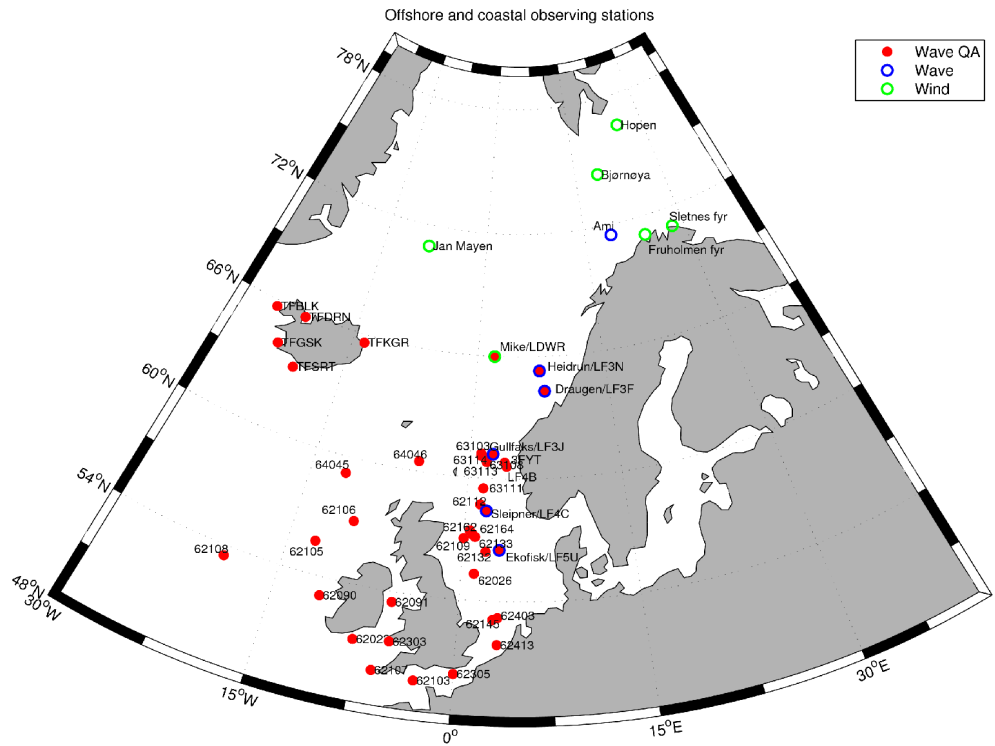

Figure 5. Coastal and offshore observing stations. In situ wave measurements covering the North-East Atlantic Ocean averaged over four hours on the synoptic times (00, 06, 12 and 18 UTC), quality-assured and prepared by ECMWF [Saetra and Bidlot, 2004] are shown in red. The measurements cover the period 1991-2002. Offshore and coastal stations only measuring wind are shown in green. Stations measuring both wind and waves from the Norwegian sector are shown in blue (partially overlapping with the quality-assured data from ECMWF). The measurements cover the period 1980-2002. Note that Weather ship M (66° N, 002° E) is marked as both "Mike" and LDWR.

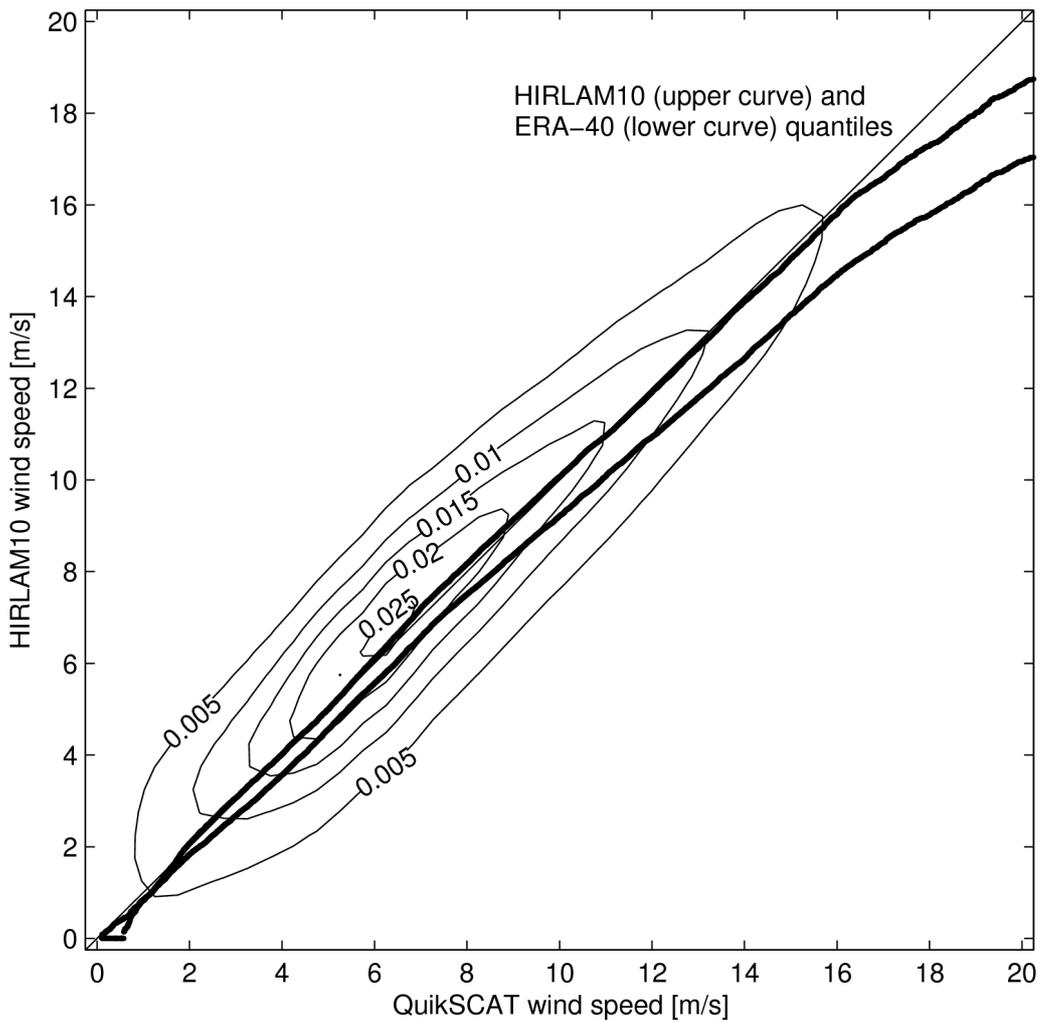

Figure 6. Joint probability density function (contoured scatter density) and quantiles v quantiles of co-located HIRLAM10 and QuikSCAT wind speed (upper curve), Jul 1999-Aug 2002. The quantiles of ERA-40 *v.* QuikSCAT are shown for reference (lower curve).

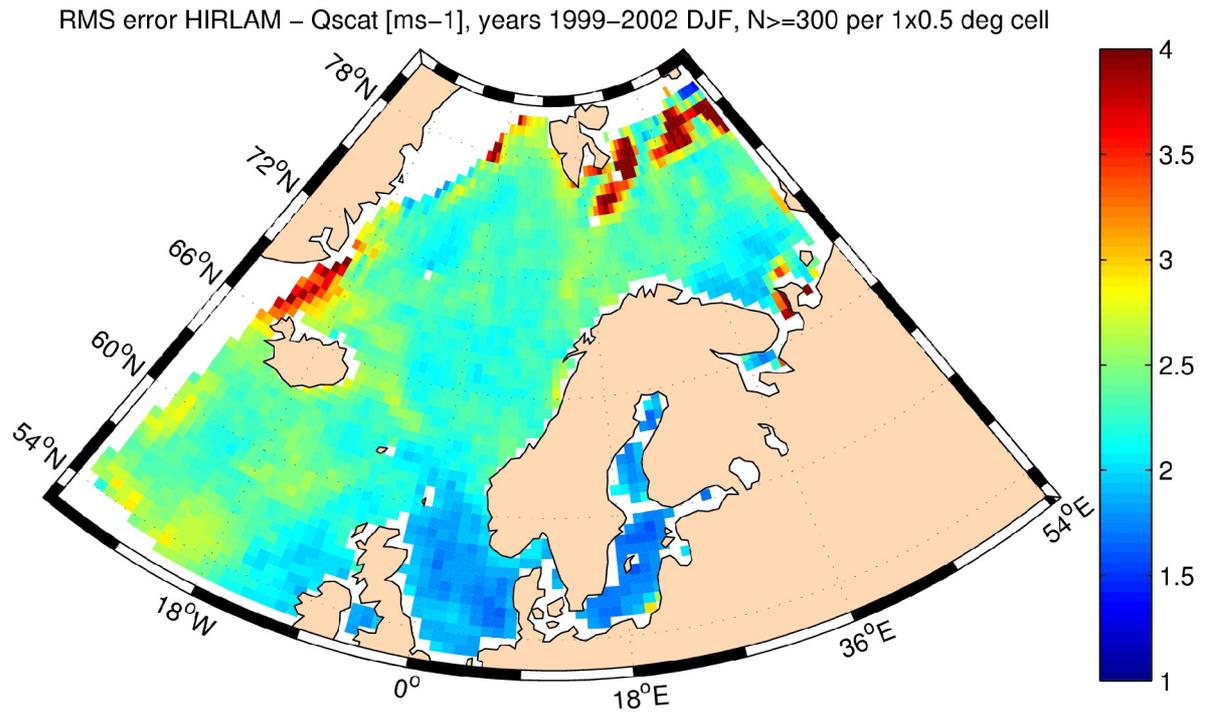

Figure 7. 10 m wind speed RMS difference between co-located HIRLAM10 and QuikSCAT, December-February, 1999-2002.

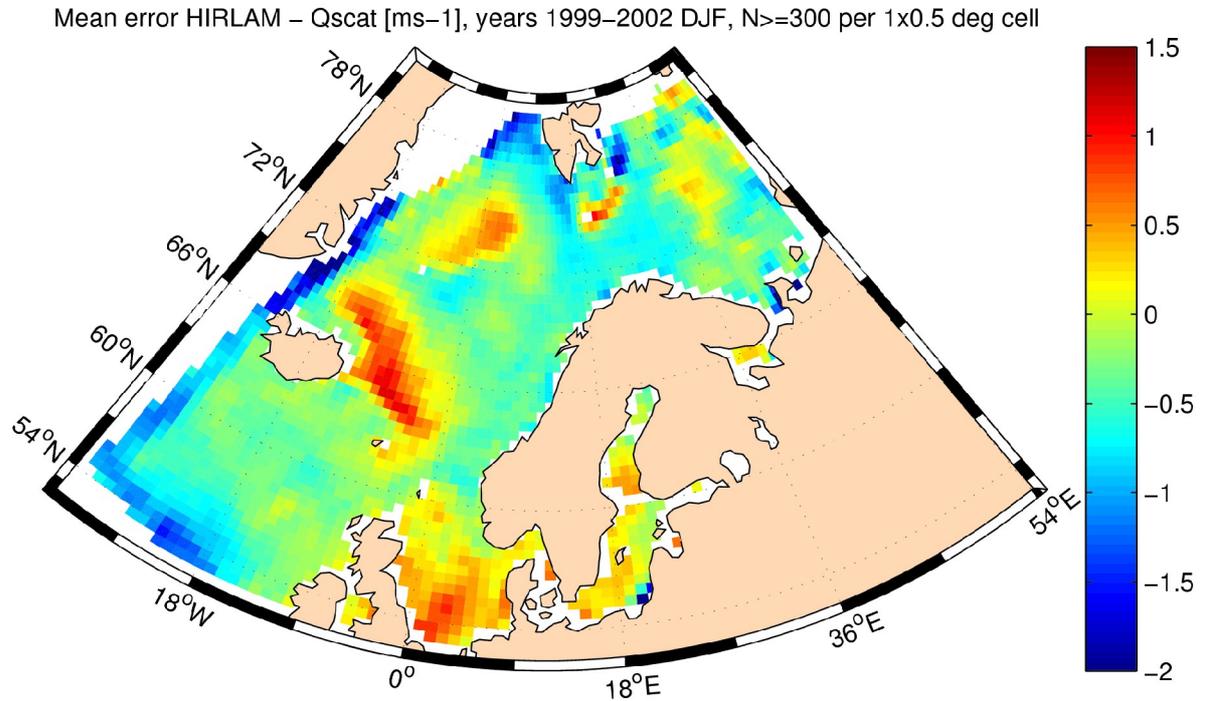

Figure 8. 10 m wind speed vector mean difference between co-located HIRLAM10 and QuikSCAT, December-February, 1999-2002.

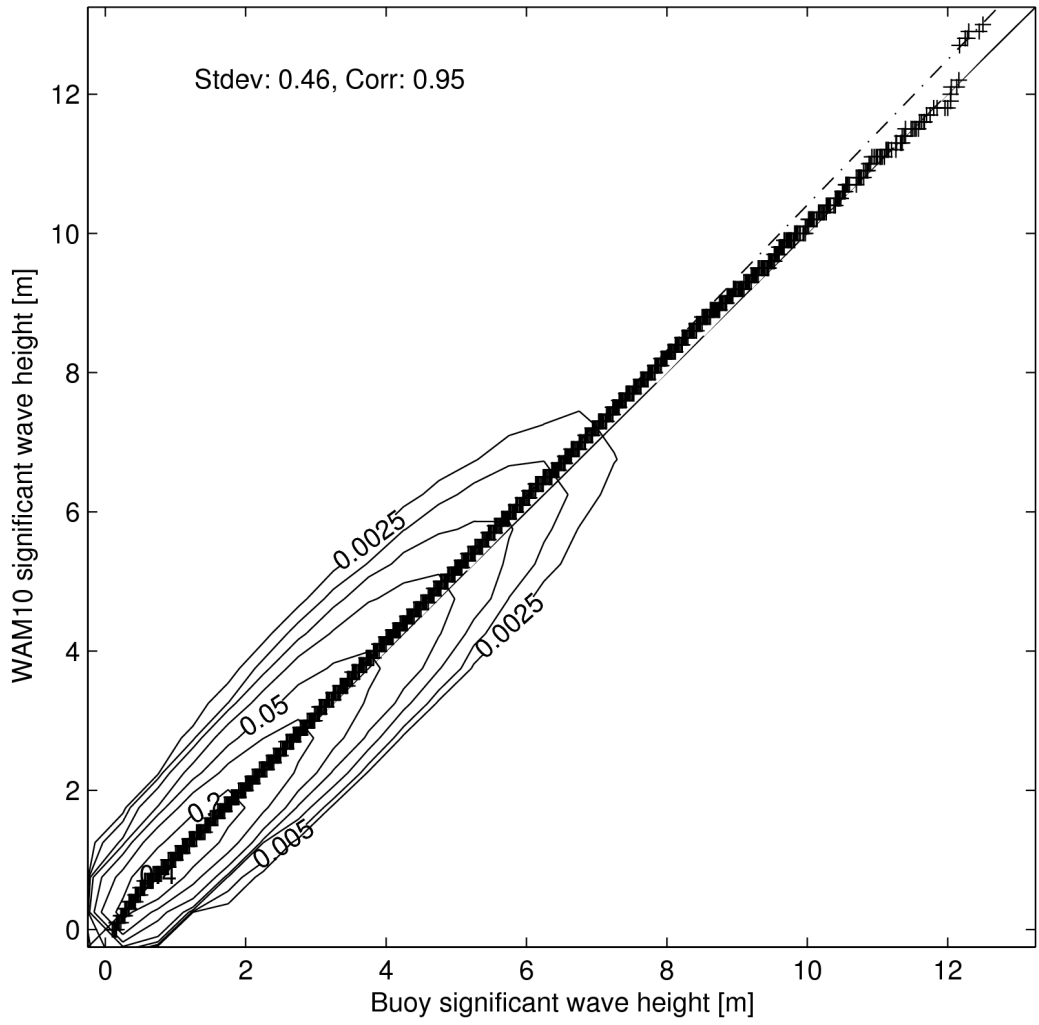

(a)

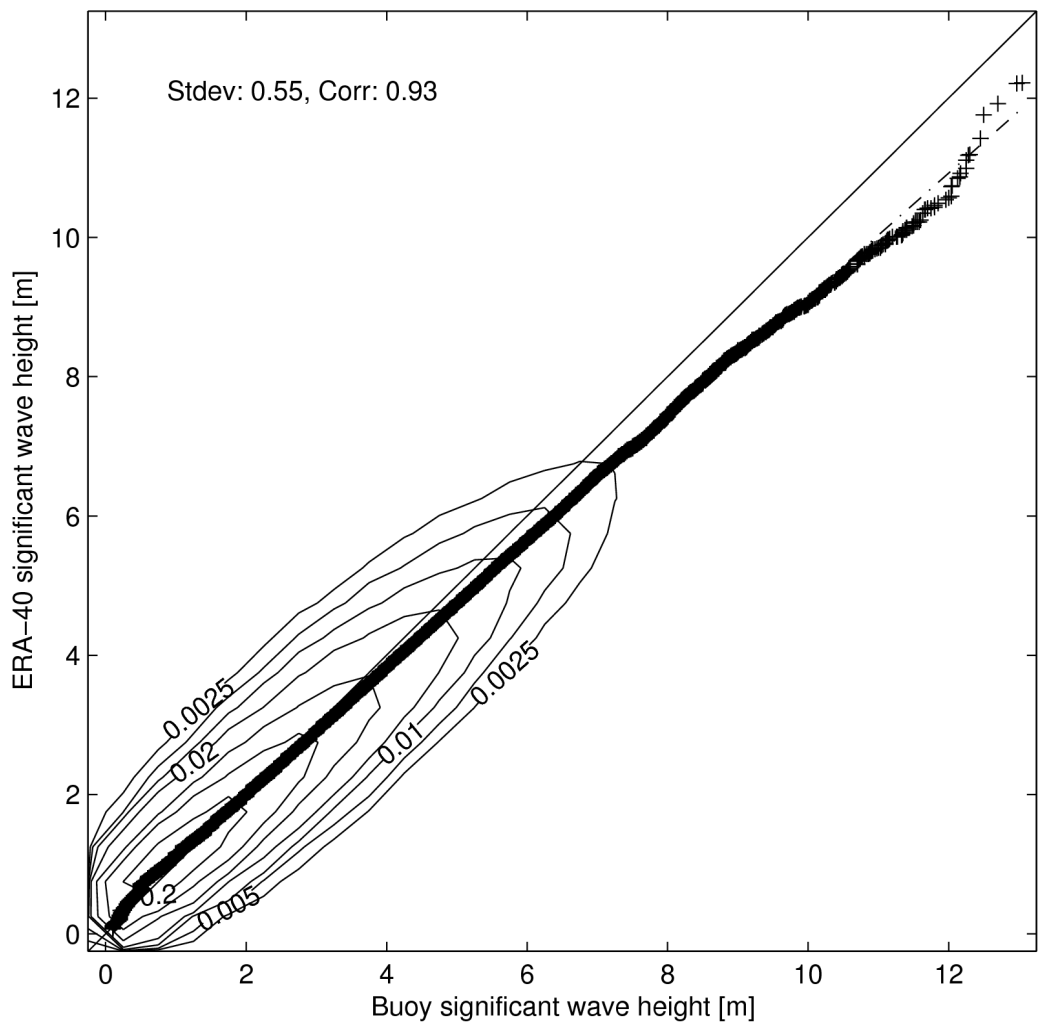

(b)

Figure 9. Scatter diagrams and qq-plots of significant wave height at 40 quality controlled buoy stations (locations shown in Figure 5), for the period 1991-2002. (a): Observations v. WAM10. (b): Observations *v.* ERA-40.

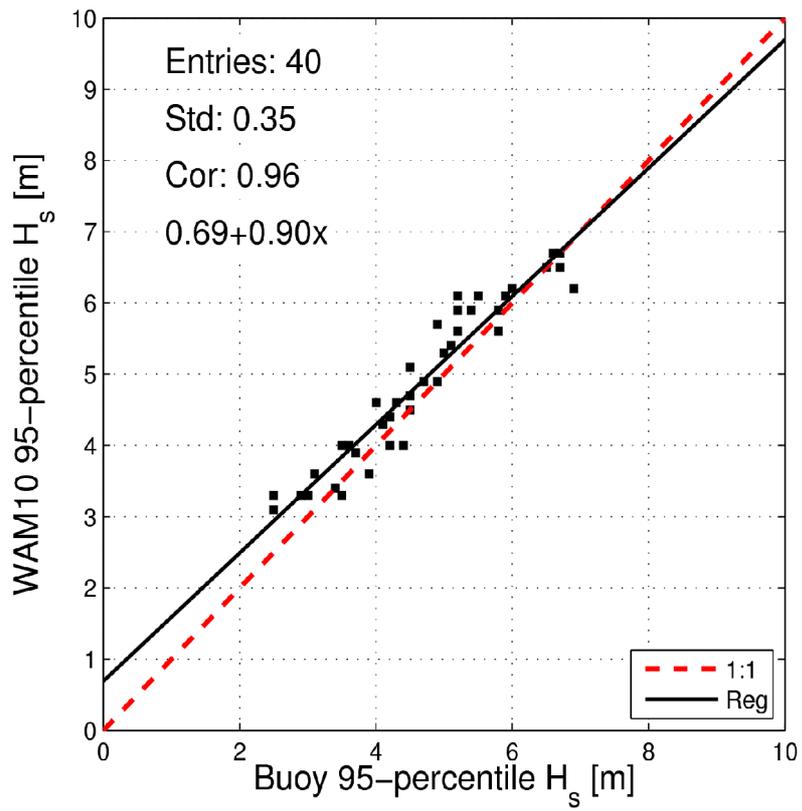

(a)

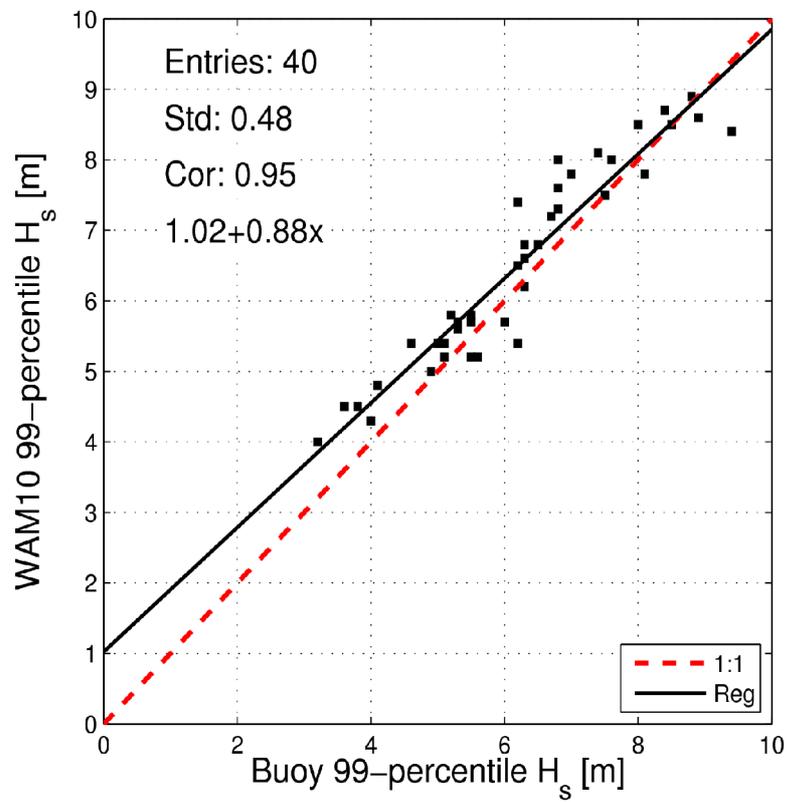

(b)

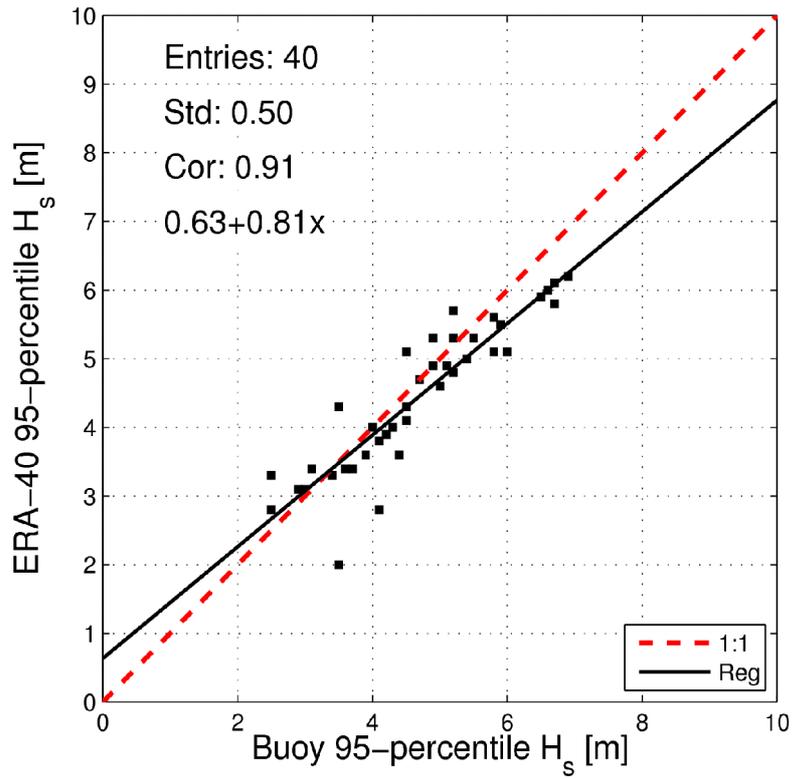

(c)

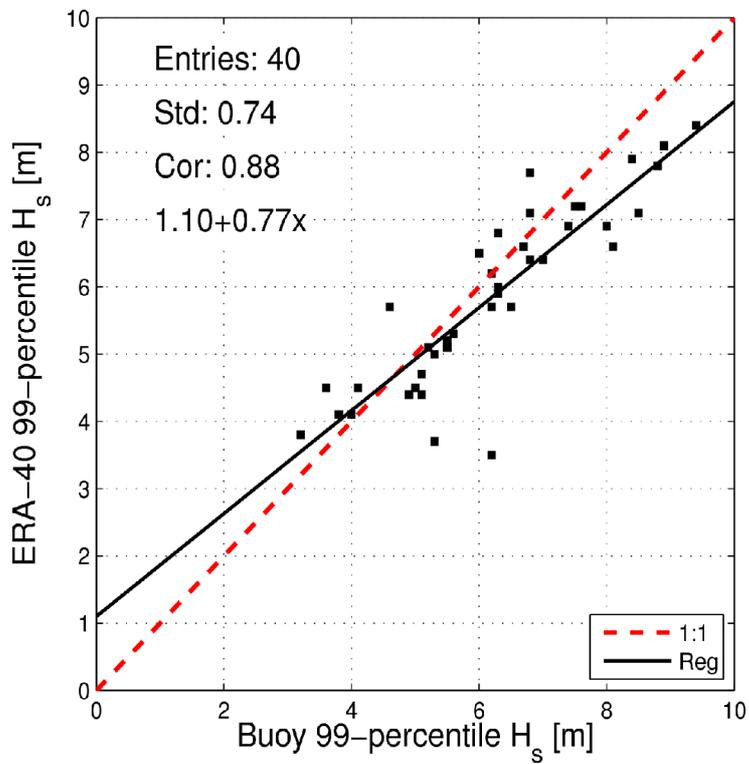

(d)

Figure 10. Modeled v observed percentiles of significant wave height at 40 quality controlled buoy stations (locations shown in Figure 5), for the period 1991-2002. WAM10: (a) 95% and (b) 99%. ERA-40: (c) 95% and (d) 99%.

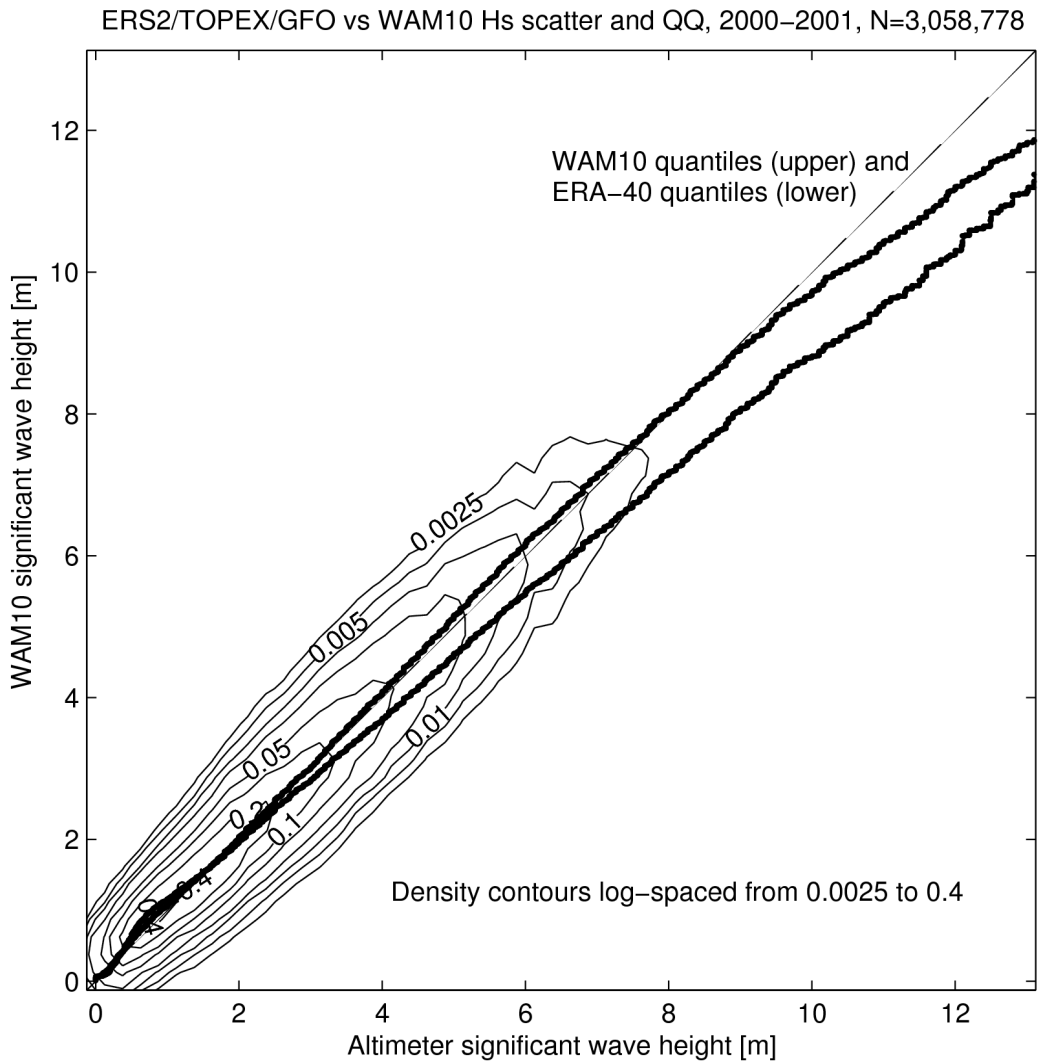

Figure 11. Joint probability density function (log-spaced contoured scatter density) and quantiles v quantiles (upper curve) of co-located ERS-2, TOPEX and GFO altimeter $H_s$ and WAM10, years 2000-2001. The quantiles of ERA-40 *v* altimeter are included for reference (lower curve).

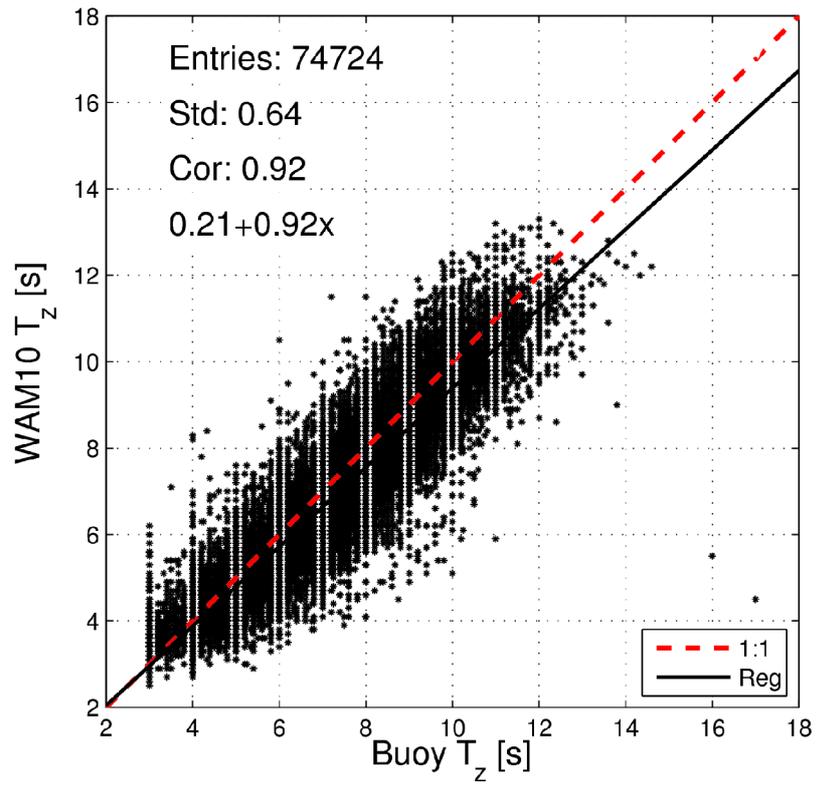

(a)

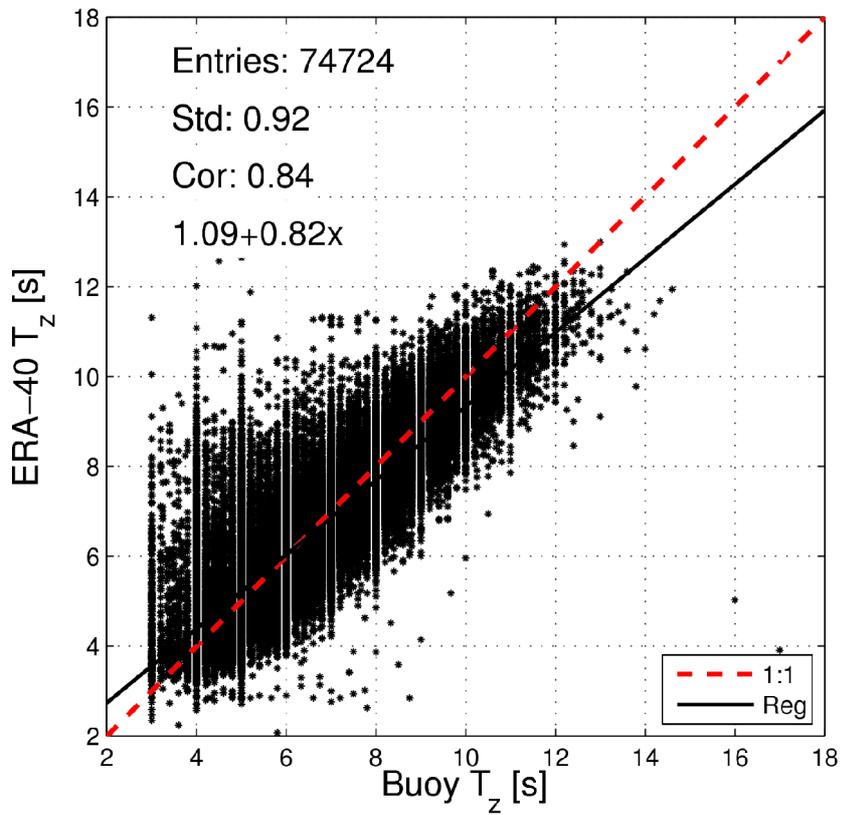

(b)

Figure 12. Observed v modeled (a: WAM10, b: ERA-40) mean period, $T_m$ [s], for the period 1991-2002. Only data from a subset of quality-controlled stations found in Figure 5 were used.

# Tables

| WIND 10 m (m s$^{-1}$) | ME | MAE | RMSE |
|---|---|---|---|
| HIRLAM10 *v* obs stations | 0.03 | 1.95 | 2.46 |
| ERA-40 *v* obs stations | -0.86 | 2.36 | 2.94 |
| HIRLAM10 *v* QuikSCAT | -0.02 | 1.50 | 2.08 |
| ERA-40 *v* QuikSCAT | -0.77 | 1.63 | 2.32 |

Table 1.

| Case | Measured | | HIRLAM10 | | ERA-40 | |
|---|---|---|---|---|---|---|
| | MSLP | U10 | MSLP | U10 | MSLP | U10 |
| 1999-12-19T13:30UTC 72ºN, 018ºE | 989 hPa | 23 m s$^{-1}$ | 992 hPa | 17 m s$^{-1}$ | 994 hPa | 14 m s$^{-1}$ |
| 2000-01-31T06:10UTC 65ºN, 004ºE | 978 hPa | 26 m s$^{-1}$ | 985 hPa | 20 m s$^{-1}$ | 988 hPa | 14 m s$^{-1}$ |
| 2001-11-01T02:00UTC 71ºN, 019ºE | 992 hPa | 26 m s$^{-1}$ | 998 hPa | 15 m s$^{-1}$ | 1000 hPa | 12 m s$^{-1}$ |

Table 2. A summary of three polar low cases and their representation in HIRLAM10 and ERA-40. The downscaling only captured the first stages of the evolution of the lows and underestimates the wind speed and the depth of the lows. ERA-40 is generally unable to model polar lows whose typical dimension is below the spectral resolution of the model (T159).

| Ekofisk 1980-2002 | | | | | | | | | |
|---|---|---|---|---|---|---|---|---|---|
| $H_s$ | N | Mean | St.dev. | Mean abs. difference | RMS difference | Corr. coefficient | P90 | P95 | P99 |
| Obs. | 58072 | 2.08 | 1.28 | - | - | - | 3.8 | 4.5 | 6.2 |
| WAM10 | 58072 | 2.08 | 1.31 | 0.28 | 0.42 | 0.95 | 3.9 | 4.6 | 6.3 |
| **Sleipner 1995-2002** | | | | | | | | | |
| Obs. | 17688 | 2.29 | 1.38 | - | - | - | 4.2 | 5.0 | 6.5 |
| WAM10 | 17688 | 2.44 | 1.38 | 0.42 | 0.57 | 0.92 | 4.4 | 5.1 | 6.7 |
| **Gullfaks 1990-2002** | | | | | | | | | |
| Obs. | 30502 | 2.68 | 1.54 | - | - | - | 4.9 | 5.7 | 7.3 |
| WAM10 | 30502 | 2.77 | 1.55 | 0.36 | 0.51 | 0.95 | 5.0 | 5.9 | 7.7 |
| **Draugen 1995-2002** | | | | | | | | | |
| Obs. | 18980 | 2.62 | 1.67 | - | - | - | 5.0 | 6.0 | 7.9 |
| WAM10 | 18980 | 2.67 | 1.64 | 0.42 | 0.58 | 0.94 | 4.9 | 6.0 | 8.2 |
| **Heidrun 1996-2002** | | | | | | | | | |
| Obs. | 16136 | 2.67 | 1.49 | - | - | - | 4.7 | 5.6 | 7.5 |
| WAM10 | 16136 | 2.73 | 1.63 | 0.48 | 0.65 | 0.92 | 5.0 | 6.1 | 8.1 |
| **Ami 1993-1998** | | | | | | | | | |
| Obs | 7462 | 2.41 | 1.45 | - | - | - | 4.3 | 5.3 | 7.5 |
| WAM10 | 7462 | 2.42 | 1.49 | 0.36 | 0.52 | 0.94 | 4.3 | 5.4 | 8.0 |

Table 3. Significant wave height observed *v.* WAM10 at the Norwegian offshore stations Ekofisk, Sleipner, Gullfaks, Draugen, Heidrun and the weather ship Ami. WAM10 is stored on three-hourly resolution. Only observations near the model time have been selected.

| Ekofisk 1980-2002 | | | | | | | | | |
|---|---|---|---|---|---|---|---|---|---|
| $H_s$ | N obs | Mean | St.dev. | Mean abs. difference | RMS difference | Corr. coefficient | P90 | P95 | P99 |
| Obs. | 29044 | 2.08 | 1.29 | - | - | - | 3.8 | 4.5 | 6.3 |
| ERA-40 | 29044 | 1.88 | 1.11 | 0.36 | 0.56 | 0.91 | 3.4 | 4.1 | 5.5 |
| **Sleipner 1995-2002** | | | | | | | | | |
| Obs. | 8889 | 2.29 | 1.38 | - | - | - | 4.2 | 4.9 | 6.4 |
| ERA-40 | 8889 | 2.19 | 1.16 | 0.45 | 0.62 | 0.90 | 3.8 | 4.5 | 5.8 |
| **Gullfaks 1990-2002** | | | | | | | | | |
| Obs. | 15230 | 2.69 | 1.55 | - | - | - | 4.9 | 5.7 | 7.4 |
| ERA-40 | 15230 | 2.60 | 1.35 | 0.46 | 0.65 | 0.91 | 4.5 | 5.3 | 6.8 |
| **Draugen 1995-2002** | | | | | | | | | |
| Obs. | 9497 | 2.62 | 1.67 | - | - | - | 5.0 | 6.0 | 7.9 |
| ERA-40 | 9497 | 2.35 | 1.27 | 0.50 | 0.70 | 0.94 | 4.1 | 4.9 | 6.6 |
| **Heidrun 1996-2002** | | | | | | | | | |
| Obs. | 8043 | 2.67 | 1.48 | - | - | - | 4.7 | 5.5 | 7.5 |
| ERA-40 | 8043 | 2.43 | 1.30 | 0.45 | 0.62 | 0.92 | 4.2 | 5.1 | 6.7 |
| **Ami 1993-1998** | | | | | | | | | |

| | | | | | | | | |
|---|---|---|---|---|---|---|---|---|
| Obs | 3730 | 2.41 | 1.45 | - | - | - | 4.3 | 5.3 | 7.4 |
| ERA-40 | 3730 | 2.07 | 1.17 | 0.44 | 0.65 | 0.93 | 3.5 | 4.3 | 6.6 |

Table 4. Significant wave height observed *v.* ERA-40 at the Norwegian offshore stations Ekofisk, Sleipner, Gullfaks, Draugen, Heidrun and the weather ship Ami. Note that as ERA-40 is only available on six-hourly resolution the number of data points, N, is roughly half of what is found in the previous table comparing WAM10 to the same observational dataset.

| $H_s$ | N obs | ME | RMSE | MAE | SI | Corr | M stn | P95 NME | P95 SI | P95 Corr | P99 NME | P99 SI | P99 Corr |
|---|---|---|---|---|---|---|---|---|---|---|---|---|---|
| WAM10 | 235368 | 0.07 m | 0.47 m | 0.33 m | 20.8 % | 0.95 | 40 | 5.0 % | 8.9 % | 0.96 | 4.7 % | 9.0 % | 0.95 |
| ERA-40 | 235368 | -0.01 m | 0.55 m | 0.40 m | 24.2 % | 0.93 | 40 | -5.0% | 11.80 % | 0.91 | -5.7 % | 13.0 % | 0.88 |

Table 5. Significant wave height observed *v.* WAM10 and ERA-40 at the 40 coastal and offshore stations [Saetra and Bidlot, 2004]. A total of 235,368 observations were recorded and quality controlled. The observations were averaged over 4 hours. The right part of the table compares the normalized mean error (NME) and the scatter index (SI) of the 95 and 99 percentiles for M=40 stations.

| $H_s$ (m) | ME | MAE | RMSE | a | b(m) | Corr |
|---|---|---|---|---|---|---|
| WAM10 *v* altimeters | 0.05 | 0.35 | 0.48 | 0.95 | 0.16 | 0.95 |
| ERA-40 *v* altimeters | -0.08 | 0.36 | 0.52 | 0.82 | 0.36 | 0.94 |

Table 6. WAM10 and ERA-40 comparison with co-located significant wave height from altimeters onboard ERS2, TOPEX and GFO (years 2000-2001). Number of observations is approximately 3,000,000. Here a and b are regression coefficients (slope and offset).

**Tables**